\documentclass[%
 preprint,
superscriptaddress, linenumbers
 amsmath,amssymb,
 aps, prd
]{revtex4-1}

\usepackage{graphicx}% Include figure files
\usepackage{dcolumn}% Align table columns on decimal point
\usepackage{bm}% bold math
\usepackage[%showframe,%Uncomment any one of the following lines to test 
margin=1.0in,
]{geometry}
\usepackage{slashed}
\usepackage{hyperref}
\usepackage{contour}
\usepackage{lmodern}
\usepackage{ulem}
\usepackage{amsmath}
\usepackage{mathrsfs}
\usepackage{multirow}
\usepackage{xcolor}
\usepackage{booktabs}
\usepackage{graphicx}
\usepackage[caption=false]{subfig}
\usepackage{feynmp-auto}
\usepackage{tcolorbox}
\usepackage{upgreek}
\usepackage{float}
\usepackage{placeins}
\usepackage[switch*]{lineno}
\newcommand{\DW}[2]{\Gamma(#1\rightarrow#2)}
\newcommand{\BR}[2]{\textrm{BR}(#1\rightarrow#2)}
\newcommand{\cW}{\tilde{c}_{W}}
\newcommand{\cZ}{\tilde{c}_{Z}}
\newcommand{\cH}{\tilde{c}_{H}}
\newcommand{\cA}{\tilde{c}_{A}}
\newcommand{\cV}{\tilde{c}_{V}}
\newcommand{\PNWA}{\textrm{P}_\textrm{NWA}(M_Q,\vec{c})}

\begin{document}
%\preprint{APS/123-QED}

\title{Novel Interpretation Strategy for Searches of Singly Produced Vector-like Quarks at the LHC}
\author{Avik Roy}
\email{aroy@utexas.edu}
\affiliation{%
 Center for Particles and Fields, Department of Physics \\
 The University of Texas at Austin, Austin, Texas 78712, USA 
}%
\author{Nikiforos Nikiforou}
\email{nikiforos.nikiforou@cern.ch}
\affiliation{%
 Center for Particles and Fields, Department of Physics \\
 The University of Texas at Austin, Austin, Texas 78712, USA 
}
\author{Nuno Castro}
\email{nuno.castro@cern.ch}
\affiliation{Laborat\'orio de Instrumenta\c{c}\~{a}o e F\'isica Experimental de Part\'iculas (LIP), \\
Universidade do Minho, 4710-057 Braga, Portugal
%LIP, Departamento de F\`isica, Escola de Ci\^encias \\
%Universidade do Minho, 4710-057 Braga, Portugal
}
\affiliation{Departamento de F\'isica, Escola de Ci\^encias, \\
Universidade do Minho, 4710-057 Braga, Portugal
%Departamento de F\`isica e Astronomia \\ 
%Faculdade de Ci\^encias, Universidade do Porto, 4169-007 Porto, Portugal
}
\author{Timothy Andeen}
\email{tandeen@utexas.edu}
\affiliation{%
 Center for Particles and Fields, Department of Physics \\
 The University of Texas at Austin, Austin, Texas 78712, USA 
}

\date{\today}% It is always \today, today,

\begin{abstract}
Vector-like Quarks (VLQs) are potential signatures of physics beyond the Standard Model at the TeV energy scale and major efforts have been put forward at both ATLAS and CMS experiments in search of these particles. In order to make these search results more relatable in the context of most plausible theories of VLQs, it is deemed important to present the analysis results in a general fashion.  We investigate the challenges associated with such interpretations of singly produced VLQ searches and propose a generalized, semi-analytical framework that allows a model-independent casting of the results in terms of unconstrained, free parameters of the VLQ Lagrangian. We also propose a simple parameterization of the correction factor to the  single VLQ production cross-section at large decay widths. We illustrate how the proposed framework can be used to conveniently represent statistical limits by numerically reinterpreting results from benchmark ATLAS and CMS analyses.

%\begin{description}
%\end{description}
\end{abstract}

\maketitle

%\tableofcontents

\section{Introduction}
Vector-like Quarks (VLQs), stringent predictions of a number of beyond Standard Model theories~\cite{new-fermion, comp-higgs, comp-higgs-2, comp-higgs-3, littlest-higgs, extra-dim, extra-dim-2, extra-dim-bounds, VL-fermion, TP-LHC} at the TeV scale, are excellent search candidates in the post-Higgs era at CERN's Large Hadron Collider (LHC). VLQs are $SU(3)$ color-triplets with the same strong coupling as the standard model quarks but maintain identical electroweak representation for both chiralities. The spectrum of the VLQ species consists of four particles, denoted as $X_{+\frac{5}{3}}, T_{+\frac{2}{3}}, B_{-\frac{1}{3}}$ and $Y_{-\frac{4}{3}}$ where the subscript indicates the electric charge of the corresponding particle. They can exist as $(T)$ or $(B)$ singlets, $(X,T)$, $(T,B)$, or $(B,Y)$ doublets and $(X,T,B)$ or $(T, B, Y)$ triplets.  In most representations, they couple to the standard model quarks via exchange of charged $(W^+, W^-)$ or neutral $(Z, H)$ bosons.

\begin{figure}
 \centering
 \subfloat[]{
  \includegraphics[width=0.5\columnwidth]{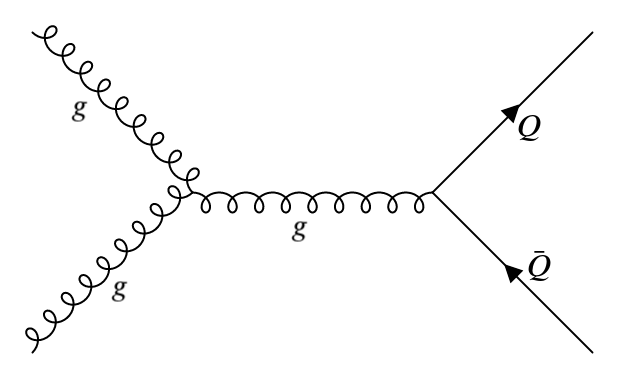}
  \label{fig:PP-diag}
  }
  \subfloat[]{
  \includegraphics[width=0.4\columnwidth]{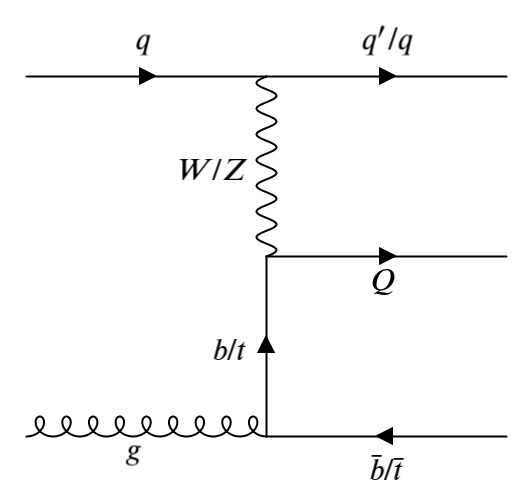}
  \label{fig:SP-diag}
  }
  \caption[]{Dominant contributing diagrams for \subref{fig:PP-diag} pair production and \subref{fig:SP-diag} single production of VLQs}
  \label{fig:PPSP-diag}
 \end{figure}

The search efforts for VLQs in collider experiments such as ATLAS and CMS can be broadly categorized into two classes: (a) seaches for VLQ pairs and (b) searches for singly produced VLQs. In general, each analysis targets a generic final state that is dominantly sensitive to one, or occasionally more than one, decay modes of the VLQs. Searches for pair production of VLQs have been traditionally more popular compared to searches for singly produced VLQs.
This is primarily because, in many theoretical models, pair production of VLQs is dominated by a model-independent, strong-force-mediated process (Figure \ref{fig:PP-diag}). However, it should be noted that alternate production modes of pair produced VLQs, e.g. via heavy gluons~\cite{HeavyGluon-1, HeavyGluon-3}, and interpretation of VLQ search results in context of such models~\cite{HeavyGluon-2} have also been explored.
%This allows a straightforward interpretation of the search results in context of most VLQ representations. 
Using the data collected at 8 TeV center of mass energy during Run 1 at the LHC between 2009 and 2013, a number of analyses concentrated on pair production of VLQs~\cite{ATLAS-run1-PP-1, ATLAS-run1-PP-2, ATLAS-run1-PP-3, CMS-run1-PP-1, CMS-run1-PP-2, CMS-run1-PP-3}. Although no significant excess was seen in data, each analysis independently set exclusion limits on a VLQ mass in the range of approximately \mbox{600--1000 GeV} that has served as the benchmark for the complementary Run 2 searches. Similar limits were obtained from the searches that focused on single production of VLQs in Run 1~\cite{ATLAS-run1-PP-3, ATLAS-TWb-run1}. 

At a center of mass energy of 13 TeV in Run 2, the ATLAS collaboration has not only performed a number of searches looking for the pair production of  VLQs~\cite{ATLAS-run2-PP-1, ATLAS-run2-PP-2, ATLAS-run2-PP-4,  ATLAS-run2-PP-5,  ATLAS-run2-PP-6} but also combined the results of these analyses to set the strongest current limits on the VLQ masses~\cite{ATLAS-PPcomb}. Complementary pair production analyses from CMS~\cite{CMS-run2-PP-1, CMS-run2-PP-2, CMS-run2-PP-3} have also set limits in the range of $\mathcal{O}(1 \textrm{ TeV})$ for up- and down-type VLQs. 

During Run 2, there has been a significant increase in the number of searches of singly produced VLQs~\cite{ATLAS-TZt-run2, ATLAS-TWb-run2, ATLAS-monotop, CMS-TZt-run2,  CMS-XWt-run2, CMS-BHb-run2, CMS-fullhad-run2}. This is partly because, depending on how strongly VLQs couple with SM bosons and quarks, single production processes can have a larger cross-section at the range of masses that Run 2 searches have been focusing on~\citep{Aguilar}. However, unlike pair production, the production of single VLQs is dominated by electroweak processes (Figure \ref{fig:SP-diag}) and they decay via exchange of electroweak gauge bosons and the Higgs boson. Hence, production and decay of single VLQs depend on the electroweak representation of these heavy fermions. 
Different analyses have adapted different strategies, often inspired by model-specific assumptions. 
Results obtained by these analyses cannot be consistently compared or combined because of the diverse set of assumptions and model-dependent interpretation strategies.

This paper aims to lay out an experimentally-inspired, semi-analytical framework for a relatively  model-independent interpretation of single VLQ production search results that can be adapted by most ongoing and future analyses. In section \ref{phil}, we explain the details of and the challenges to a model-independent interpretation of single VLQ production searches and emphasize on why such a strategy is important. Section \ref{framework} introduces a minimal set of assumptions and presents the semi-analytical framework. In section \ref{PNWA}, we introduce a novel parameterization for estimating the correction to the single VLQ production cross-section at finite widths. Finally, in section \ref{recast}, we demonstrate how this framework can be used to compare, reinterpret and visualize existing search results from ATLAS and CMS.

\section{The Philosophy of Interpretation of VLQ Searches}\label{phil}

A standard experimental search for VLQs benefits from a relatively \textit{model-independent} parameterization of these particles. Such a representation utilizes a collection of arbitrary parameters- the VLQ masses $\{M_Q\}$ and their couplings $\vec{c} = \{{c}_{L/R, V/H}^{Qq}\}$ to the Standard Model quarks via exchange of the gauge bosons, $V \in \{W^\pm, Z\}$ and the Higgs boson, $H$. An experimental search evaluates the statistically excluded cross-section, $\sigma_{\textrm{lim}} \left( M_Q, \vec{c} \right)$, for a grid of points of the parametric hyperspace.  These limits can be  interpreted in the context of a certain theoretical model as long as the model does not drammatically deviate from the assumptions of the model-independent representation. This approach has been dubbed the ``Bridge Model'' by Matsedonsky et. al~\cite{Wulzer}. In the same paper, the authors describe the following simplified Lagrangian  for VLQs in terms of these generalized couplings,
\begin{equation}\label{SimpLag}
\mathcal{L} = \sum_{\zeta, q, Q} \left[ \frac{g_w}{2} \sum_{V} c_{\zeta,V}^{Qq}\bar{Q}_\zeta \slashed{V} q_\zeta + c_{\zeta, H}^{Qq} H \bar{Q}_{\zeta '} q_\zeta \right] + \mathrm{h.c.}
\end{equation}
where $Q$ represents the usual VLQs $\{X_{+\frac{5}{3}}, T_{+\frac{2}{3}}, B_{-\frac{1}{3}}, Y_{-\frac{4}{3}}\}$, $\zeta$ and $\zeta '$ represent alternate chiralities and $q$ represents a SM quark of up or down type. Some of these couplings may be constrained by the conservation of certain quantum numbers. For example, the $+\frac{5}{3}$ charged partner $X$ can only couple with SM up-type quarks by the exchange of a $W$ boson. The parametic hyperspace of this model-independent representation can be mapped to those of similar formulations in~\cite{Panizzi, Fuks, Atre, Harada, Aguilar} by a one-to-one correspondence among the tree-level couplings.  

\begin{figure*}
 \centering
 \subfloat[]{
  \includegraphics[width=0.5\textwidth]{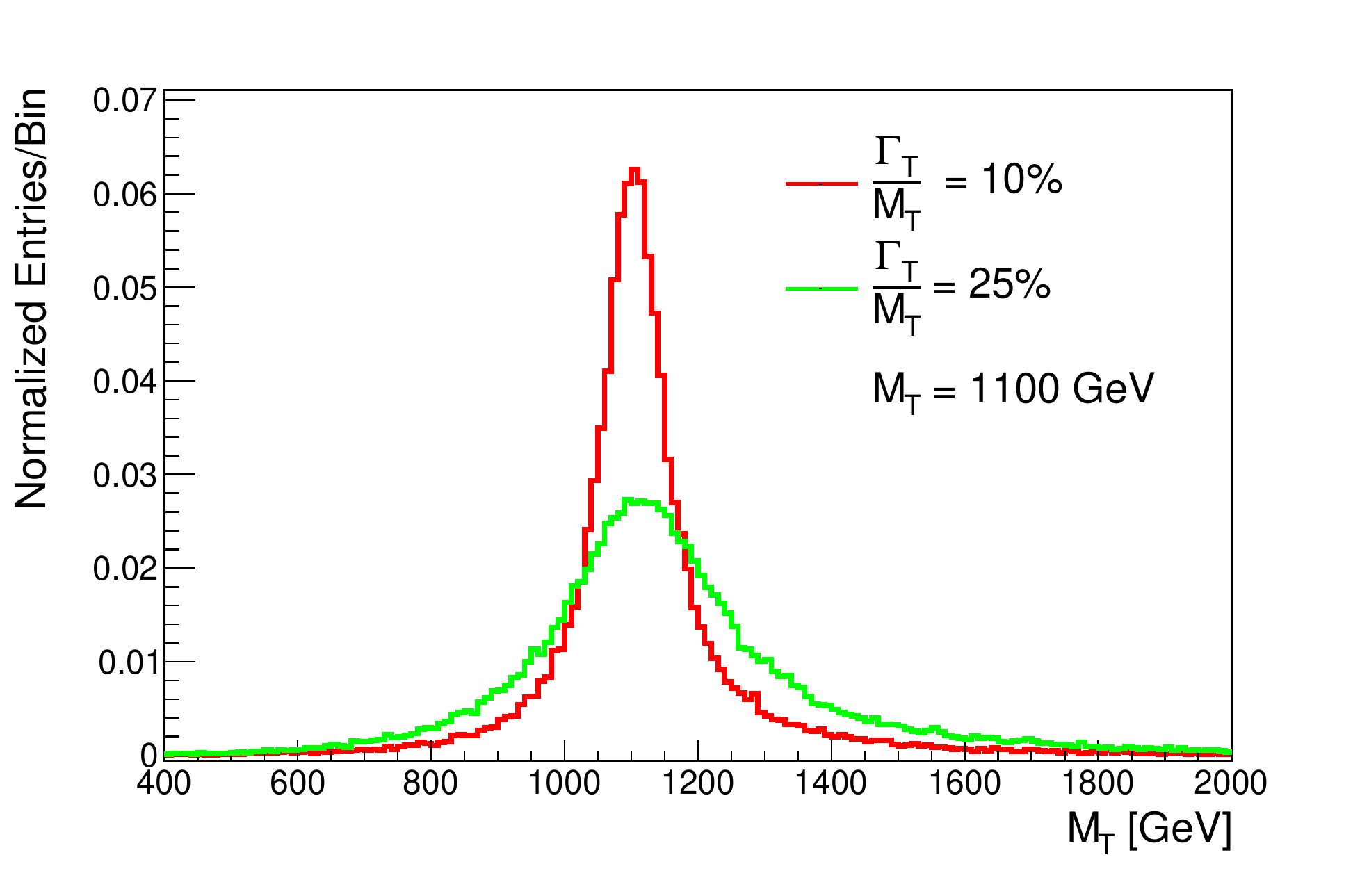}
  \label{fig:M_VLQ}
  }
  \subfloat[]{
  \includegraphics[width=0.5\textwidth]{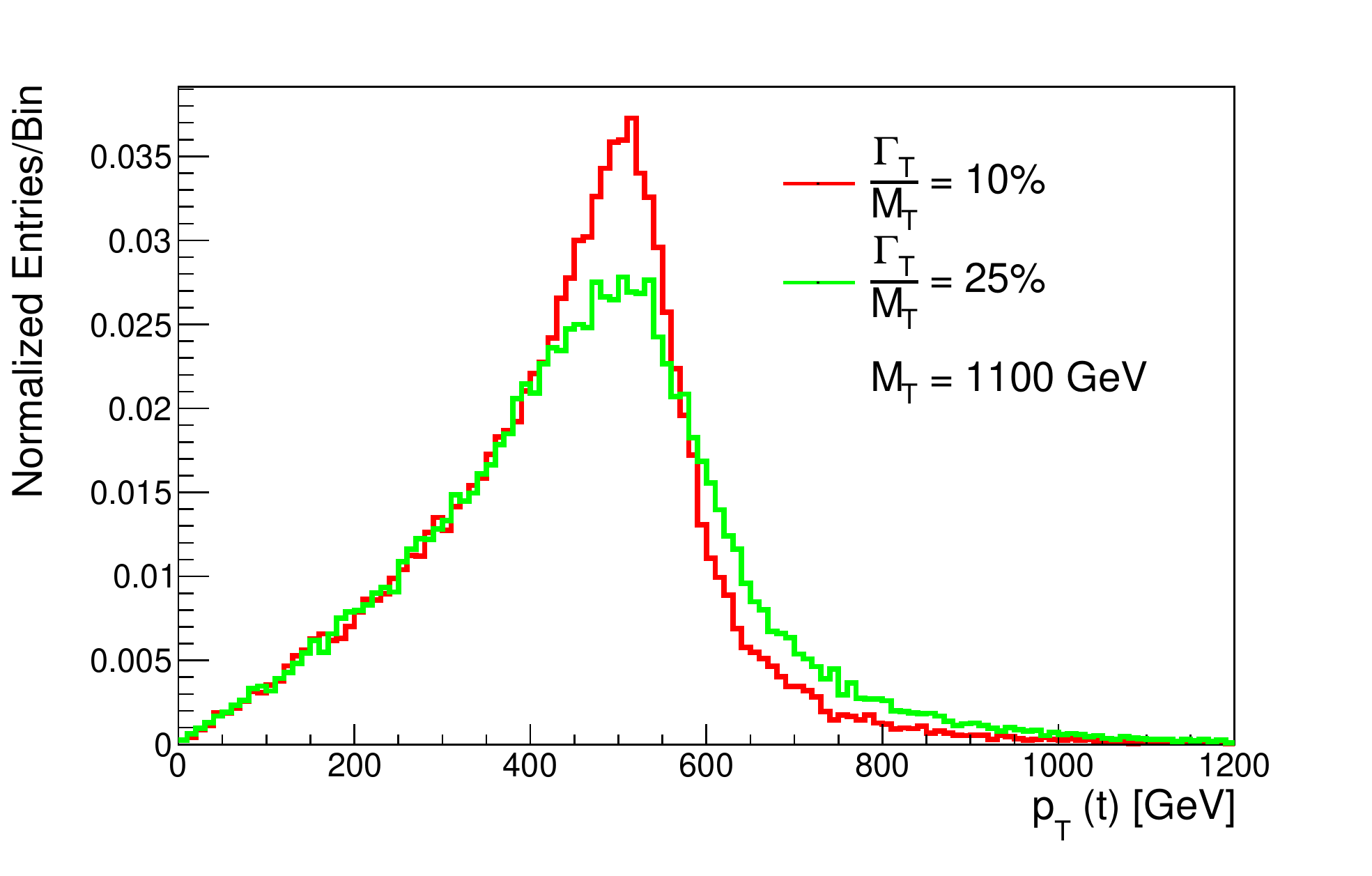}
  \label{fig:pT_top}
  }
  \caption[]{Distribution of \subref{fig:M_VLQ} VLQ invariant mass and \subref{fig:pT_top} transverse momentum of the top quark from its decay in association with a $Z$ boson for a $+\frac{2}{3}$ charged Top partner ($T$). These distributions were obtained from Monte-Carlo simulation of singly produced $T$ using \textsc{MadGraph\_aMC@NLO}~\cite{MG-1,MG-2}. The coupling values are set to $c_{L,W} = 0.5$ (red) and $c_{L,W} = 0.8$ (green) with branching ratios  assumed to be $50\%, 25\%$ and $25\%$ for the $Wb, Zt$ and $Ht$ decay modes, respectively, to obtain the desired relative decay width. All histograms are normalized to unity.}
  \label{fig:kin_VLQ}
 \end{figure*}

Pair production of VLQs has inspired an elegant interpretation strategy. The dominant, QCD-facilitated production mode for a pair of VLQs allows a model-independent estimate of the cross section for the pair production under a narrow width approximation (NWA),~i.e.:
\small{
\begin{widetext}
\begin{align}
\sigma(pp \rightarrow Q\bar{Q} \rightarrow V_1q_1V_2\bar{q}_2; M_Q, \vec{c}) \enspace  \xrightarrow[]{\textrm{NWA}}  \enspace  \sigma_{\textrm{prod, } Q\bar{Q}}^{\textrm{NW}}(M_Q) \times \textrm{BR}(Q \rightarrow V_1q_1; \vec{c}) \times \textrm{BR}(Q \rightarrow V_2q_2; \vec{c})
\end{align}
\end{widetext}
}
Here, the value of the production cross-section $\sigma_{\textrm{prod, } Q\bar{Q}}^{\textrm{NW}}(M_Q)$ is independent of the electroweak group representation and the corresponding coupling parameters of the heavy fermions. This allows for a reformulation of the VLQ Lagrangian by treating the branching ratios themselves as free parameters while ignoring their complex dependence on the coupling parameters, $\vec{c}$. Moreover, most analyses looking for these heavy fermionic resonances also assume that these particles couple predominantly only with the third generation of standard model quarks, i.e.:
\begin{equation}\label{eqn:BRassumption}
\BR{Q}{Hq} + \BR{Q}{Zq} + \BR{Q}{Wq'} = 1.0
\end{equation}
where $q,q' \in \{t,b\}$. 

The assumption on the branching ratios in equation (\ref{eqn:BRassumption}) allows, along with the NWA, for a simple interpretation for searches of pair production of VLQs, where the excluded region in the parametric hyperspace is evaluated by solving the following inequality:
\begin{equation}\label{eqn:PP-interp}
\sigma_{\textrm{prod,  } Q\bar{Q}}^{\textrm{NW}} \left ( M_Q \right ) \ge \sigma_{\textrm{lim,  } Q\bar{Q}}^{\textrm{NW}} \left( M_Q, \textrm{BR}_W, \textrm{BR}_H \right )
\end{equation}
where $\sigma_{\textrm{lim,  } Q\bar{Q}}^{\textrm{NW}}$ is the statistically excluded cross section limit (usually computed at 95\% confidence level) at narrow width, corresponding to the largest process cross-section compatible with the background only hypothesis given the observed distribution in data. The cross-section limit depends on the sensitivity of an analysis to different VLQ decay modes and hence, is a function of the choice of VLQ decay branching fractions. As a result, a parametrically model-independent interpretation of pair production searches can be done by solving equation (\ref{eqn:PP-interp}) for some chosen grid of allowed values of $\BR{Q}{Wq'}$, $\BR{Q}{Hq}$ to set a limit on the VLQ mass.

The importance of a universal, model-independent interpretation strategy is pivotal for a combination of multiple analyses. A consistent combination of multiple analyses requires a well defined correlation scheme among the various nuisance parameters as well as the parameter(s) of interest (POI), the latter usually being a function of signal cross-section. A generalized interpretation strategy, as given in equation~(\ref{eqn:PP-interp}), can be used to formulate a well-defined correlation scheme for the POIs of different analyses. Different analyses tend to be sensitive to different kinematic signatures and different regions of the phase space. A combination of such analyses guided by a well defined interpretation strategy can significantly boost the statistical power and hence, set stronger limits on the parametric hyperspace. The combination~\cite{ATLAS-PPcomb} performed by the ATLAS collaboration, excluded up (down)-type VLQ masess up to 1.31 (1.03) TeV for any combination of branching ratios respecting equation~(\ref{eqn:BRassumption}). 

In light of the interpretation strategy for pair produced VLQs as given in equation~(\ref{eqn:PP-interp}), a framework of interpreting search results for singly produced VLQs can be laid out. The excluded region of parametric hyperspace can be evaluated by solving the inequality
\begin{equation}\label{eqn:SP-interp-0} 
\sigma_{VQAq} \left ( M_Q, \vec{c} \right ) \ge \sigma_{\textrm{lim,  } VQAq} \left( M_Q, \vec{c} \right )
\end{equation}
where $VQAq$ is short-hand notation for the production of the VLQ $Q$ being mediated by the vector boson $V$ that subsequently decays to the boson $A$ and SM quark $q$. The production and decay of single VLQs  involve the relevant couplings at the corresponding vertices. The same couplings determine the partial decay widths and hence, the branching ratios of these VLQs in the associated decay channels. Moreover, the kinematic distributions of VLQ decay products also change with the change of VLQ decay widths and hence, with the choice of couplings (Figure \ref{fig:kin_VLQ}). This changes the phase space sensitivity of the analyses and, as a result, the exclusion limit also becomes non-trivially dependent on the choice of not only the VLQ mass but also the couplings.

The non-trivial coupling dependence of the interpretation relation in equation (\ref{eqn:SP-interp-0}) makes it somewhat challenging to formulate a generalized interpretation strategy. Previous ATLAS and CMS analyses incorporated simplified interpretation strategies by making model-dependent assumptions either about the branching ratios~\cite{ATLAS-TWb-run1, ATLAS-TWb-run2} or about the relative decay widths~\cite{CMS-XWt-run2, CMS-TZt-run2}, or by interpreting the generic couplings in terms of the mixing angles between the heavy fermions and their SM counterparts~\cite{ATLAS-TWb-run2, ATLAS-TZt-run2, ATLAS-monotop}. Such diversity of interpretation strategies often makes it difficult to compare results from different analyses and formulate a consistent correlation scheme for a possible combination of such analyses. 

In the following section, we present the formulation of a generalized interpretation strategy for singly produced VLQs. Following the footsteps of the interpretation strategy of searches for pair production of VLQs, we pursue a general idea of making a set of strategic assumptions to reduce the dimensions of the parametric hyperspace providing the avenue of translating the exclusion limits to excluded regions of the reduced hyperspace. 

\section{Framework for Interpretation of Singly Produced VLQs}\label{framework}

The \textit{first assumption} we make is to restrict VLQs to interact with third generation SM quarks exclusively. Additionally, we exclude models that incorporate single production of VLQs in association with heavy vector bosons~\cite{NewVB-1,NewVB-2} or their decay via exotic scalars~\cite{exot-1, exot-2, exot-3, exot-4, exot-5, exot-6} although such models are often theoretically well-motivated. This results in the branching fractions to be constrained according to equation (\ref{eqn:BRassumption}). In light of this assumption, we will drop the superscript indices $Qq$ in the generalized couplings from this point forward-- this association will be clear from the context of discussion.

\textit{Secondly,} we assume that VLQs are much heavier than the SM fermions and bosons, i.e.: 
\begin{equation}\label{eqn:mass-assumption}
M_Q \gg m_t
\end{equation}
where $m_t$ denotes the top quark mass of \mbox{$172.5$ GeV}~\cite{PDG}. This assumption is inspired by the model-independent pair production search results from the ATLAS and CMS collaborations, which set a limit on VLQ masses in the range of $\mathcal{O}(\textrm{1 TeV})$ independently of its electroweak representation.  At the large $M_Q$ limit, the interference term between left and right handed couplings in the analytic expressions of the partial decay widths become negligible. Therefore, the corresponding partial decay widths for $Q \rightarrow Vq$ and $Q \rightarrow Hq$ can be approximated as:
\begin{widetext}
 \begin{align}
  \DW{Q}{Vq} &\approx \left(c_{L,V}^2 + c_{R,V}^2\right)\times\frac{g_w^2}{32\pi}\frac{p(M_Q,m_q,m_V)}{M_Q^2}\left(\frac{M_Q^2 + m_q^2}{2} + \frac{(M_Q^2 - m_q^2)^2}{2m_V^2} - m_V^2\right) \label{eqn:gamma_V} \\
    \DW{Q}{Hq} &\approx \left(c_{L,H}^2 + c_{R,H}^2\right)\times\frac{1}{8\pi}\frac{p(M_Q,m_q,m_H)}{M_Q^2}\frac{M_Q^2 + m_q^2 - m_H^2}{2} \label{eqn:gamma_H}
 \end{align}
 \end{widetext}
where $g_w$ represents the electroweak coupling constant and,
\begin{equation*}
 p(X,y,z) = \frac{1}{2X}\sqrt{[X^2-(y+z)^2][X^2-(y-z)^2]} .
\end{equation*}
 
Since the decay width expressions only depend on the quadrature sum of the left and right handed couplings, we introduce a more convenient set of notations:
 
 \begin{equation}\label{eqn:LR}
  c_{W/Z/H}^2  = c_{L, W/Z/H}^2 + c_{R, W/Z/H}^2 .
 \end{equation}

Defining $r_A = \frac{m_A}{M_Q}$, we introduce the following functions:
\begin{widetext}
\begin{align}\label{eqn:rhos}
 \rho_W(Q) & = \sqrt{1+r_W^4+r_q^4-2r_W^2-2r_q^2-2r_W^2r_q^2} (1+r_W^2-2r_q^2-2r_W^4+r_q^4+r_W^2r_q^2) \nonumber \\
 \rho_Z(Q) & = \sqrt{1+r_Z^4+r_q^4-2r_Z^2-2r_q^2-2r_Z^2r_q^2} (1+r_Z^2-2r_q^2-2r_Z^4+r_q^4+r_Z^2r_q^2)\\
 \rho_H(Q) & = \sqrt{1+r_H^4+r_q^4-2r_H^2-2r_q^2-2r_H^2r_q^2} (1+r_q^2 - r_H^2) . \nonumber
\end{align}
\end{widetext}
These functions evaluate to unity at leading order, i.e. $\rho_W \approx \rho_Z \approx \rho_H \approx 1.0$ because $r \ll 1$. They introduce some minor mass-dependent corrections that vanish for large values of $M_Q$. In order to simplify the expression of the branching ratios, we rescale the coupling parameters as follows:

\begin{align}\label{eqn:rescale}
 c_W^2 & = c_{L,W}^2 + c_{R,W}^2  = \cW^2 \nonumber \\
 c_Z^2 & = c_{L,Z}^2 + c_{R,Z}^2  = \cZ^2\frac{m_Z^2}{m_W^2} \\
 c_H^2 & = c_{L,H}^2 + c_{R,H}^2  = \frac{g_w^2}{4} \cH^2 \frac{M_Q^2}{m_W^2} . \nonumber
\end{align}
Expressed in terms of the rescaled couplings and the $\rho_A(Q)$ functions, the decay widths become:

\begin{align}\label{eqn:new_DWs}
 \DW{Q}{Aq} & = \cA^2 \times \frac{g_w^2}{128\pi}\frac{M_Q^3}{m_W^2} \times \rho_A(Q) .
\end{align}
Then, the equation for branching ratio reduces to:

\begin{equation}\label{eqn:new_BRs}
 \BR{Q}{Aq} = \frac{\cA^2 \rho_A(Q)}{\cW^2 \rho_W(Q) + \cZ^2 \rho_Z(Q) + \cH^2 \rho_H(Q)} .
\end{equation}  
Hence, the assumption in equation (\ref{eqn:mass-assumption}) allows the branching fractions to be independent of the chirality of the couplings.  The same assumption makes the production cross-section of single VLQs independent of the choice of the chirality at leading order. For example, as argued by Matsedonsky et. al,~\cite{Wulzer} the production cross-section for $Z$ and $W$ boson mediated production modes under the narrow width approximation, are approximated as:

\begin{align}
\sigma_{\textrm{prod, }VQ}^{\textrm{NW}} (M_Q,\vec{c}) \approx & \left( c_{V}^2 + k \times c_{L,V}c_{R,V} \frac{m_q}{m_q + M_Q} \right) \nonumber \\
 & \times \sigma_{\textrm{prod, }VQ}^{\textrm{NW}}(M_Q, c_V = 1) \label{eqn:prod}
\end{align}
where $k$ is a constant of $\mathcal{O}(1)$. Assuming $M_Q \gg m_q$ as in equation (\ref{eqn:mass-assumption}) suppresses the interfering term by $\mathcal{O}\left( \frac{m_q}{M_Q} \right)$. Furthermore, Aguilar-Saavedra et. al proved in~\cite{Aguilar} that independent of the representation, either of the chiral couplings is suppressed by an additional factor of $\mathcal{O} \left( \frac{m_q}{M_Q} \right)$ that emerges from the diagonalization of the mass matrix. As a result, the contribution of the interference term in equation (\ref{eqn:prod}) will be of subleading order and can be ignored. Based on similar arguments, it was also assumed in~\cite{Panizzi} that either of the chiralities dominates the model-independent representation of the VLQs.
%{\color{red}
Many analsyes, in both ATLAS and CMS collaborations, have also reported that thier results are independent of the chiral structure of the couplings- the effect of different chiralities is indistinguishable within a coarse binning structure in the discriminant variables as well as the statistical and systematic uncertainties that dominate the limit setting.~\cite{ATLAS-TZt-run2, ATLAS-monotop, CMS-TZt-run2, CMS-BHb-run2, CMS-fullhad-run2} 
%}
Hence, we introduce our \textit{third assumption}; an analysis is either insensitive to the relative structure of the chiral couplings or a single chirality dominates the signal kinematics. 
As a result, the statistical limits obtained from an analysis, $\sigma_{\textrm{lim}}$, now depends on $M_Q$ and the chirality-ignorant rescaled couplings in equation (\ref{eqn:rescale}), i.e. $\vec{c} = \{\cW, \cZ, \cH \}$. 

The three assumptions made so far can be summarized as follows:

\begin{itemize}
\item \textit{VLQs as Top- or Bottom-partners:} VLQs predominantly couple to the third generation SM quarks via exchange of $W,Z$  and $H$ bosons.
\item \textit{Heavy VLQs:} VLQs are much heavier than SM bosons and fermions.
\item \textit{Chirality-agnostic Analysis:} The analysis for search of single VLQ is either insensitive to relative chiral structure or dominated by a single chirality of the couplings.
\end{itemize}

Given the aforementioned set of assumptions, we can now derive the explicit expression for the inequality in (\ref{eqn:SP-interp-0}). We can write the $VQAq$ process cross-section as a product of the production cross-section and the corresponding branching ratio for narrow widths:
\begin{widetext}
\begin{align}\label{eqn:XSprod-NW}
\sigma_{VQAq}\left(M_Q, \vec{c} \right) \quad \xrightarrow[]{\textrm{NWA}} & \quad   \sigma_{\textrm{prod, } VQ}^{\textrm{NW}}\left(M_Q, \vec{c} \right) \times \BR{Q}{Aq} \nonumber \\
& = \cV^2 \times \sigma_{\textrm{prod, } VQ}^{\textrm{NW}} \left(M_Q, \cV = 1 \right) \times \BR{Q}{Aq}
\end{align}
\end{widetext}
where the branching ratio is given by equation (\ref{eqn:new_BRs}). For larger widths, this estimate for the cross-section is corrected for the width dependence of the process cross-section. Following the recipe of~\cite{ATLAS-TWb-run2, ATLAS-TZt-run2}, we define the correction factor as:
 
\begin{equation}\label{eqn:PNWA-def}
\PNWA = \frac{\sigma_{\textrm{prod, } VQ}^{\textrm{NW}} \times \BR{Q}{Aq}}{\sigma_{VQAq}} .
\end{equation}
 
Together with equations (\ref{eqn:XSprod-NW}) and (\ref{eqn:PNWA-def}), the interpretation relation in equation (\ref{eqn:SP-interp-0}) reduces to:
 \begin{widetext}
\begin{equation}\label{eqn:SP-interp-1}
\frac{ \cV^2 \times \sigma_{\textrm{prod, } VQ}^{\textrm{NW}} \left(M_Q, \cV = 1 \right) \times \BR{Q}{Aq}}{\PNWA} \ge \sigma_{\textrm{lim,  } VQAq} \left( M_Q, \vec{c} \right ) .
\end{equation}
\end{widetext}
 
\section{Evaluation of $\textrm{P}_{\textrm{NWA}}$}\label{PNWA}

In the limit of narrow decay width, the Breit-Wigner distribution can be approximated as a delta function:
\begin{equation}\label{BW-approx}
\frac{1}{(p^2 - M^2)^2 + \Gamma^2M^2} \xrightarrow[]{\frac{\Gamma}{M} \rightarrow 0} \frac{\pi}{M\Gamma}\delta(p^2 - M^2) .
\end{equation}
For large widths, this approximation breaks down and the cross-section estimate accumulates corrections in higher order of $\frac{\Gamma}{M}$~\cite{Kauer-NWA, Kauer-NWA-2}. It is reasonable to assume that the correction factor for cross-section will also depend only on the values of $\Gamma_Q$ and $M_Q$ and not on the individual choices for the couplings. As a result, the analytic expression for the correction factor takes the following form:
\begin{equation}\label{eqn:NWA}
\PNWA \equiv \textrm{P}_{\textrm{NWA}}\left(\frac{\Gamma_Q}{M_Q}\right) \approx 1 + \sum_n A_n \left(\frac{\Gamma_Q}{M_Q}\right)^n
\end{equation}
where the values $A_n$ will depend on the choice of the VLQ mass and the process of interest.

The traditional parameterization of the VLQ Lagrangian, as given in equation (\ref{SimpLag}) and its equivalent formulations, allow signal event generation by fixing the coupling parameters while the partial decay widths can be calculated from equation (\ref{eqn:new_DWs}). Therefore, using the VLQ UFO model inspired by the parameterization presented in~\cite{Fuks}, single production of top and bottom partners for  both $W$ and $Z$ boson mediated single-$T$ and single-$B$ processes for $M_Q$ in the range of \mbox{1000--2200 GeV} in steps of \mbox{200 GeV} was simulated in \textsc{MadGraph\_aMC@NLO}. The actual choice of coupling values in relation to the representative set of couplings $C = \{ 0.05, 0.2, 0.4, 0.6, 0.8, 1.0\}$ is summarized in Table \ref{tab:couplings}. We additionally required that \mbox{$\frac{\Gamma_Q}{M_Q} < 0.5 $} for any choice of couplings and disregard any coupling combination that violates this constraint. The $\kappa, \hat{\kappa}, \tilde{\kappa}$ parameters introduced in~\cite{Fuks} can be calculated from the rescaled couplings from a one-to-one correspondence with the tree-level couplings in equation (\ref{SimpLag}).

\begin{table}[]
\centering
 \begin{tabular}{|c | c |} 
 \hline
 Process & Coupling Choice \\
 \hline
 \hline
 $WTWb, WTZt, ZTWb,$ & $\cW, \cZ \in C$\\
 $WBWt, WBZb, ZBWt$  & $\cH = 0 $ or, $\cH = \cZ$ \\
 \hline
 $WTHt,$ & $\cW, \cH \in C$ \\
 $WBHb $ & $\cZ = 0 $ or, $\cZ = \cH$ \\
 \hline
 $ZTZt, ZTHt,$ & $\cZ, \cH \in C$ \\
 $ZBZb, ZBHb$ & $\cW = 0 $ or, $\cW = \cH$ \\
 \hline
 \end{tabular}
 \caption{The choice of couplings for event generation and cross-section calculation in \textsc{MadGraph\_aMC@NLO}}
\label{tab:couplings}
\end{table}
 
We calculated the narrow-width and large-width cross section at leading order (LO) for all the processes mentioned in Table \ref{tab:couplings}. Using equation (\ref{eqn:PNWA-def}), we evaluated  the correction factor for each choice of coupling and mass.

\begin{figure*}
 \centering
  \includegraphics[width=0.80\textwidth]{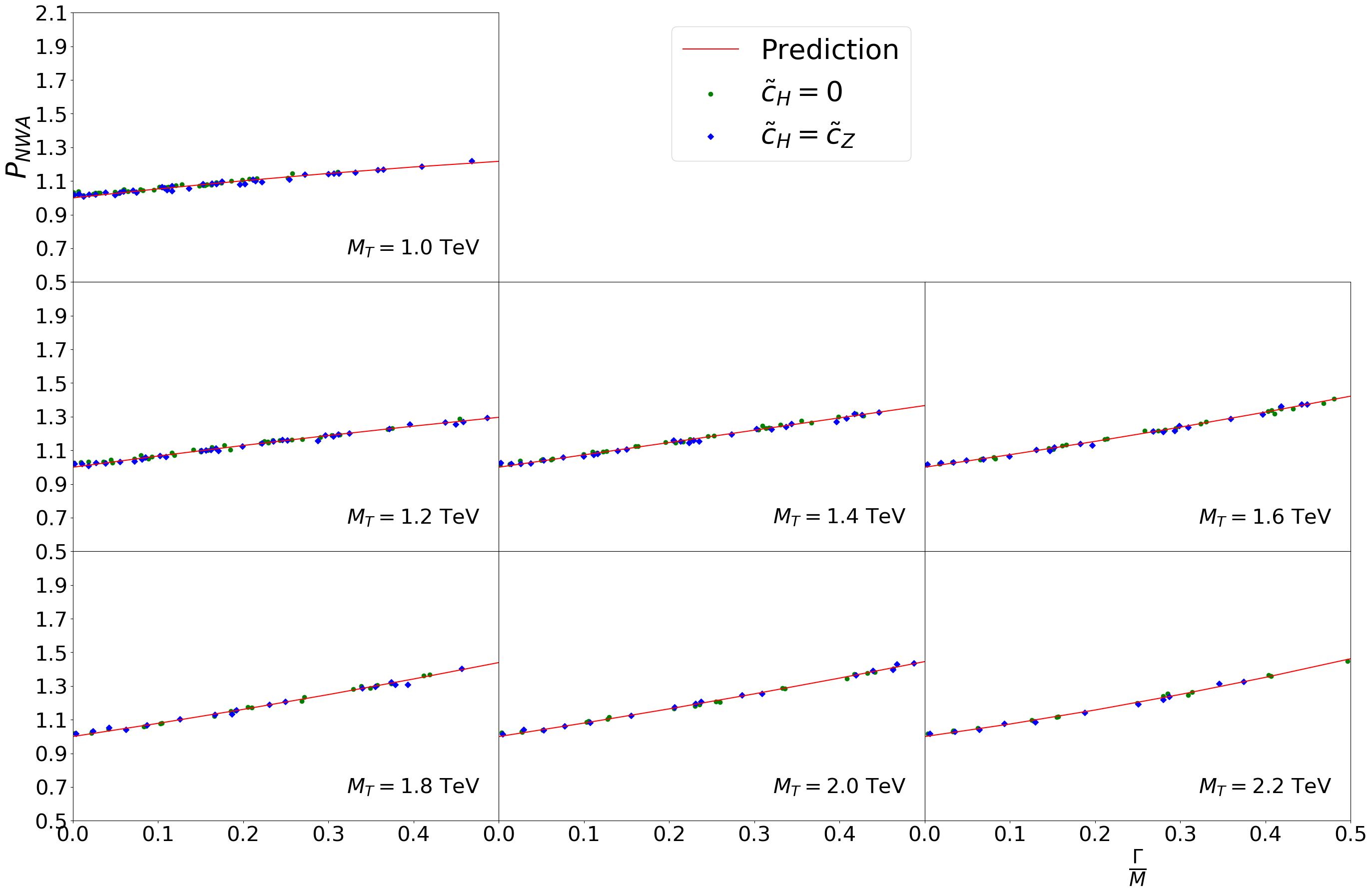}
  \caption[]{Estimated values of $\textrm{P}_{\textrm{NWA}}$ plotted as a function of $\frac{\Gamma}{M}$ for different values of $M_T$ for the $WTZt$ process. The red line shows the best fitted polynomial estimate for the correction factor.}
  \label{fig:NWA-Gamma-WTZt}
 \end{figure*}

\begin{figure*}
 \centering
  \includegraphics[width=0.80\textwidth]{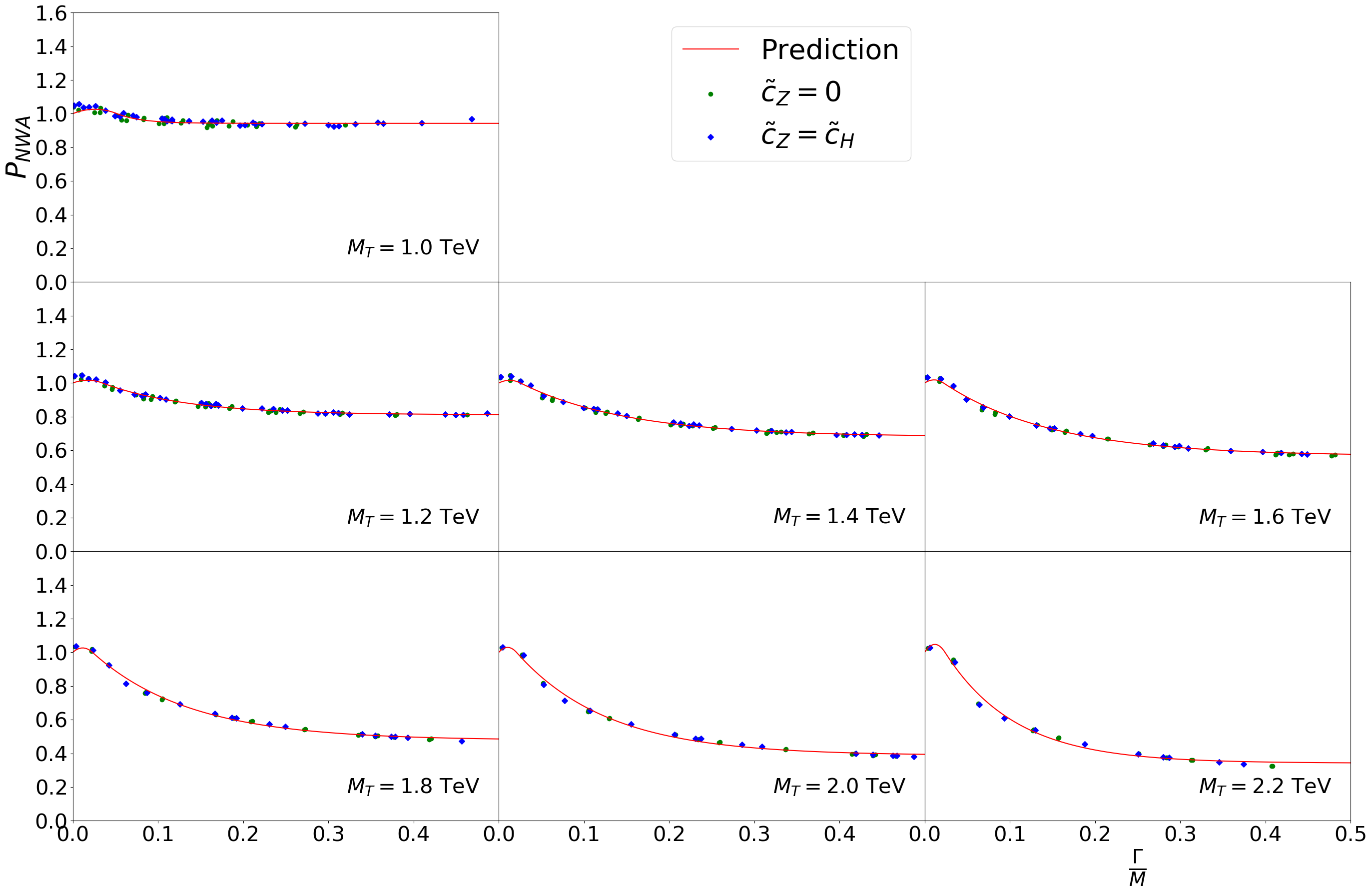}
  \caption[]{Estimated values of $\textrm{P}_{\textrm{NWA}}$ plotted as a function of $\frac{\Gamma}{M}$ for different values of $M_T$ for the $WTHt$ process. The red line shows the best fitted estimate of an exponential function for the correction factor.}
  \label{fig:NWA-Gamma-WTHt}
 \end{figure*}

Figures \ref{fig:NWA-Gamma-WTZt} and \ref{fig:NWA-Gamma-WTHt}, respectively  show the variation in $\textrm{P}_{\textrm{NWA}}$ as a function of $\frac{\Gamma}{M}$ for the $WTZt$ and $WTHt$ processes. Confirming the initial assumption, there is no strong dependence on the choice of couplings. For example, as can be seen from Figure \ref{fig:NWA-Gamma-WTZt}, for a given value of $\frac{\Gamma_Q}{M_Q}$, the  correction factor calculated for \mbox{$\cH = 0$} is almost identical to the one calculated for \mbox{$\cH = \cZ$}. 

%{\color{red}

\begin{figure*}
 \centering
 \subfloat[]{
  \includegraphics[width=0.33\textwidth]{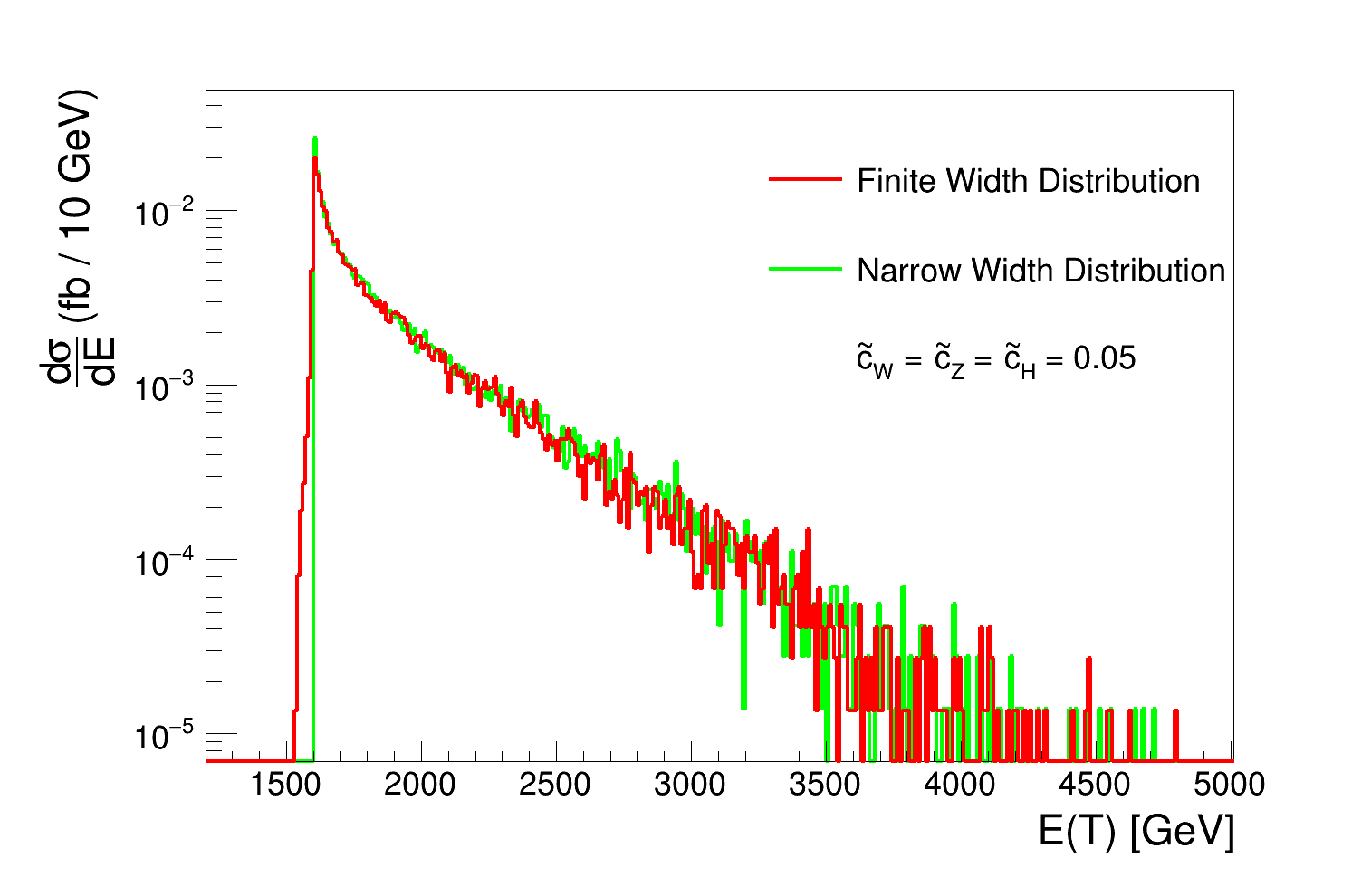}
  \label{fig:ETCompZ_005}
  }
  \subfloat[]{
  \includegraphics[width=0.33\textwidth]{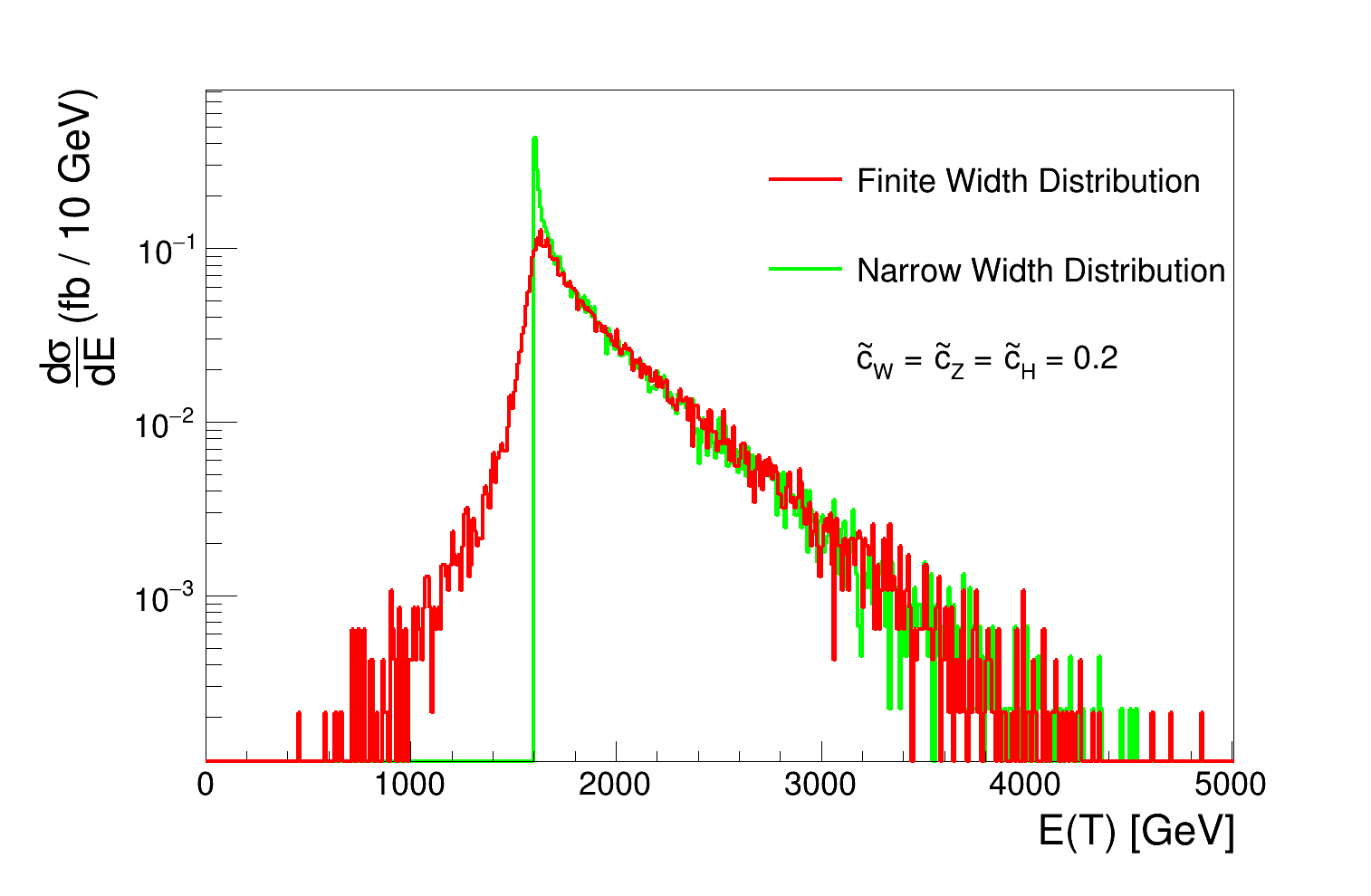}
  \label{fig:ETCompZ_020}
  }
  \subfloat[]{
  \includegraphics[width=0.33\textwidth]{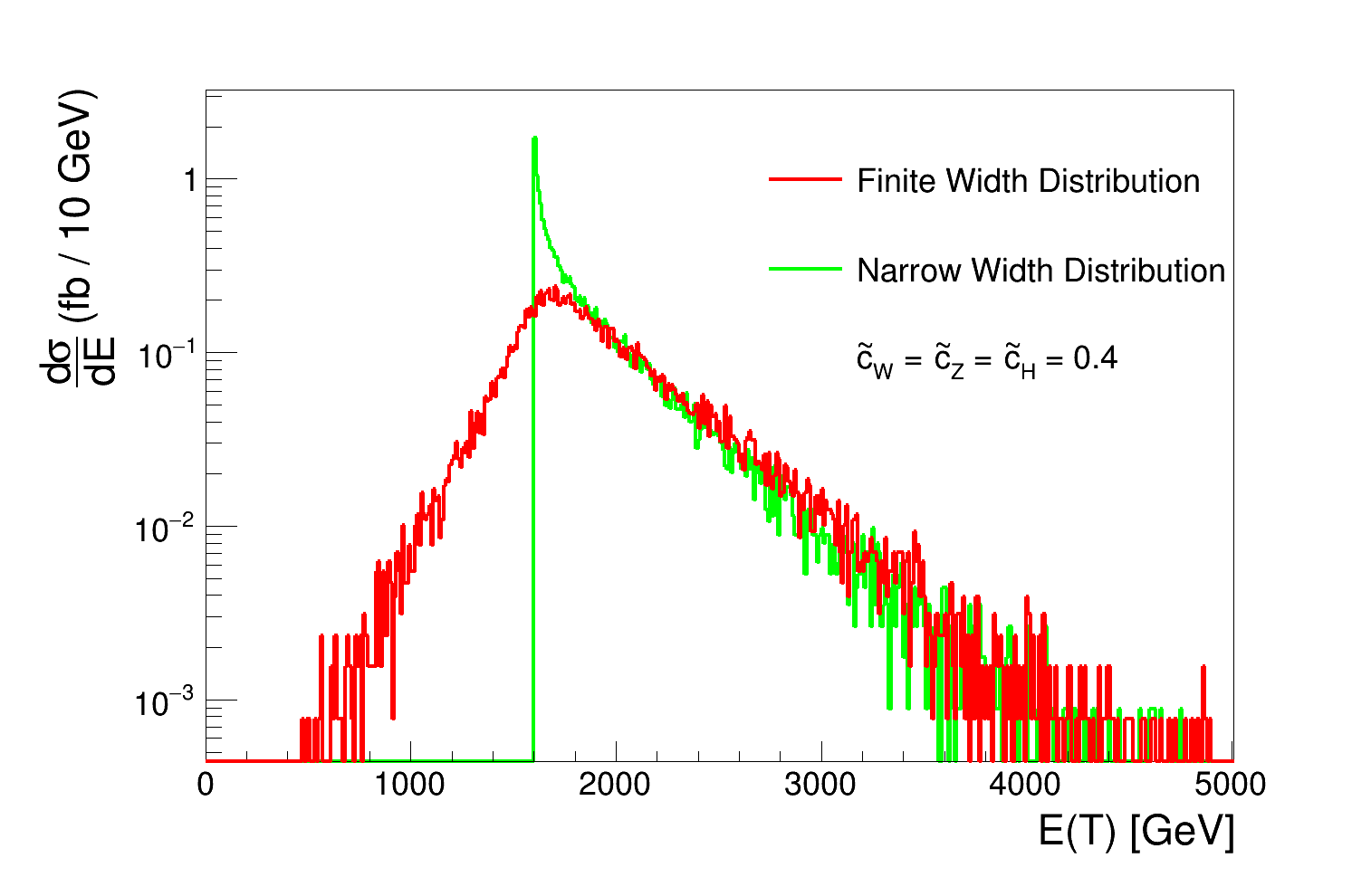}
  \label{fig:ETCompZ_040}
  }
  \caption[]{Distribution of differential cross-section of singly produced Top partner ($T$) of mass $M_T = 1.6$~TeV at narrow-width and finite-width as obtained from \textsc{MadGraph\_aMC@NLO} simulation of $WTZt$ process. The coupling values $\cW$ is set 0.05, 0.20, and 0.40 respectively in \subref{fig:ETCompZ_005}, \subref{fig:ETCompZ_020}, and \subref{fig:ETCompZ_040}. All distributions assume $\cW = \cZ = \cH$.}
  \label{fig:ETCompZ}
 \end{figure*}

We also observe in Figure \ref{fig:NWA-Gamma-WTZt} that $\textrm{P}_{\textrm{NWA}}$ monotonically rises from unity in the case of $WTZt$ processes. The differential cross-section distributions for $WTZt$ processes of a 1.6~TeV Top partner for different choices of the VLQ couplings are shown in Figure \ref{fig:ETCompZ}. Accounting for increased width causes a decrease in total cross-section because as the VLQ kinematics reach a phase space away from the pole mass, the matrix element receives a compensating contribution from the VLQ propagator. The functional behavior of  $\textrm{P}_{\textrm{NWA}}$ is well approximated by a quadratic polynomial for processes that incorporate a decay to the vector bosons. 

\begin{equation}
\label{eqn:PNWA-VQVq}
\textrm{P}_{\textrm{NWA}, VQVq} \left(\frac{\Gamma_Q}{M_Q} \right) = 1 + A_1\frac{\Gamma_Q}{M_Q} + A_2\left(\frac{\Gamma_Q}{M_Q}\right)^2
\end{equation}

The $A_1$ and $A_2$ parameters in equation (\ref{eqn:PNWA-VQVq}) for different processes with Top and Bottom partners decaying to vector bosons are evaluated by obtaining the least squared error fit to the observed values of the correction factor from simulation and  are tabulated in Tables \ref{tab:A1A2-T} and \ref{tab:A1A2-B}, respectively.

\begin{table*}[]
\centering
 \begin{tabular}{|c | c | c | c | c | c | c | c | c |}
 % c | c | c | c |} 
 \hline
   & \multicolumn{2}{|c|}{$WTWb$} 
   & \multicolumn{2}{|c|}{$WTZt$} 
%   & \multicolumn{2}{|c|}{$WTHt$} 
   & \multicolumn{2}{|c|}{$ZTWb$} 
   & \multicolumn{2}{|c|}{$ZTZt$} 
%   & \multicolumn{2}{|c|}{$ZTHt$} 
   \\
  \hline
 $M_T$ (TeV) & $A_1$ & $A_2$ 
 			 & $A_1$ & $A_2$ 
 	%		 & $A$ & $B$ 
 			 & $A_1$ & $A_2$ 
 			 & $A_1$ & $A_2$ 
 	%		 & $A$ & $B$  
 	\\ 
 \hline\hline
1.0 & 0.526 & -0.123 
	& 0.557 & -0.247 
%	& 0.080 & 5.436 
	& 0.507 	& -0.272 
	& 0.374 	& 0.013 
%	& $\sim$ 0 & $\sim$ 0
\\
1.2 & 0.638 & -0.048 
	& 0.681 & -0.176
%	& 0.221 & 5.190 
	& 0.639 & -0.126 
	& 0.550 & 0.035 
%	& 0.089 & 7.108
\\
1.4 & 0.708 & 0.054 
	& 0.733 & -0.001
%	& 0.358 & 5.043 
	& 0.737 & -0.046 
	& 0.664 & 0.087 
%	& 0.226 & 5.379
\\
1.6 & 0.697 & 0.171 
	& 0.715 & 0.257
%	& 0.468 & 5.472 
	& 0.757 & 0.141 
	& 0.730 & 0.166 
%	& 0.352 & 5.233
\\
1.8 & 0.784 & 0.153 
	& 0.759 & 0.240
%	& 0.571 & 5.819 
	& 0.807 & 0.160 
	& 0.746 & 0.312 
%	& 0.467 & 5.492
\\
2.0 & 0.764 & 0.235 
	& 0.777 & 0.227 
%	& 0.646 & 6.667 
	& 0.820 & 0.195 
	& 0.786 & 0.275 
%	& 0.560 & 6.036
\\
2.2 & 0.720 & 0.346 
	& 0.696 & 0.456 
%	& 0.715 & 7.390 
	& 0.754 & 0.356 
	& 0.709 & 0.477 
%	& 0.643 & 6.725
\\
\hline
\end{tabular}
\caption{The best fit values for the parametric representation of $\textrm{P}_{\textrm{NWA}}$ in equation (\ref{eqn:PNWA-VQVq}) for different values of $M_T$.}
\label{tab:A1A2-T}
\end{table*}

\begin{table*}[]
\centering
 \begin{tabular}{|c | c | c | c | c | c | c | c | c | }
 %c | c | c | c |} 
 \hline
   & \multicolumn{2}{|c|}{$WBWt$} 
   & \multicolumn{2}{|c|}{$WBZb$} 
%   & \multicolumn{2}{|c|}{$WBHb$} 
   & \multicolumn{2}{|c|}{$ZBWt$} 
   & \multicolumn{2}{|c|}{$ZBZb$} 
 %  & \multicolumn{2}{|c|}{$ZBHb$} 
 \\
  \hline
 $M_T$ (TeV) & $A_1$ & $A_2$ 
 			 & $A_1$ & $A_2$ 
% 			 & $A$ & $B$ 
 			 & $A_1$ & $A_2$ 
 			 & $A_1$ & $A_2$ 
 %			 & $A$ & $B$  
 \\ 
 \hline\hline
1.0 & 0.506 & -0.333 
	& 0.506 & -0.280 
%	& 0.047 & 5.136 
	& 0.548 & -0.250 
	& 0.409 & 0.133 
%	& 0.231 & 3.612
\\
1.2 & 0.652 & -0.178 
	& 0.646 & -0.159 
%	& 0.169 & 5.046
	& 0.653 & -0.111 
	& 0.542 & 0.105 
%	& 0.347 & 4.512
\\
1.4 & 0.741 & -0.046 
	& 0.721 & -0.015 
%	& 0.305 & 4.724
	& 0.727 & -0.009 
	& 0.625 & 0.166 
%	& 0.465 & 4.870
\\
1.6 & 0.776 & 0.116 
	& 0.759 & 0.133 
%	& 0.417 & 5.098 
	& 0.761 & 0.108 
	& 0.677 & 0.210 
%	& 0.555 & 5.597
\\
1.8 & 0.807 & 0.195 
	& 0.796 & 0.189 
%	& 0.519 & 5.565 
	& 0.760 & 0.222
	& 0.673 & 0.351 
%	& 0.638 & 6.266
\\
2.0 & 0.831 & 0.203 
	& 0.806 & 0.235 
%	& 0.606 & 6.093 
	& 0.768 & 0.242 
	& 0.711 & 0.300 
%	& 0.707 & 6.960
\\
2.2 & 0.781 & 0.345 
	& 0.757 & 0.346 
%	& 0.679 & 6.955 
	& 0.720 & 0.362 
	& 0.648 & 0.442 
%	& 0.763 & 8.004
\\
\hline
\end{tabular}
\caption{The best fit values for the parametric representation of $\textrm{P}_{\textrm{NWA}}$ in equation (\ref{eqn:PNWA-VQVq}) for different values of $M_B$.}
\label{tab:A1A2-B}
\end{table*}

\begin{figure*}
 \centering
 \subfloat[]{
  \includegraphics[width=0.33\textwidth]{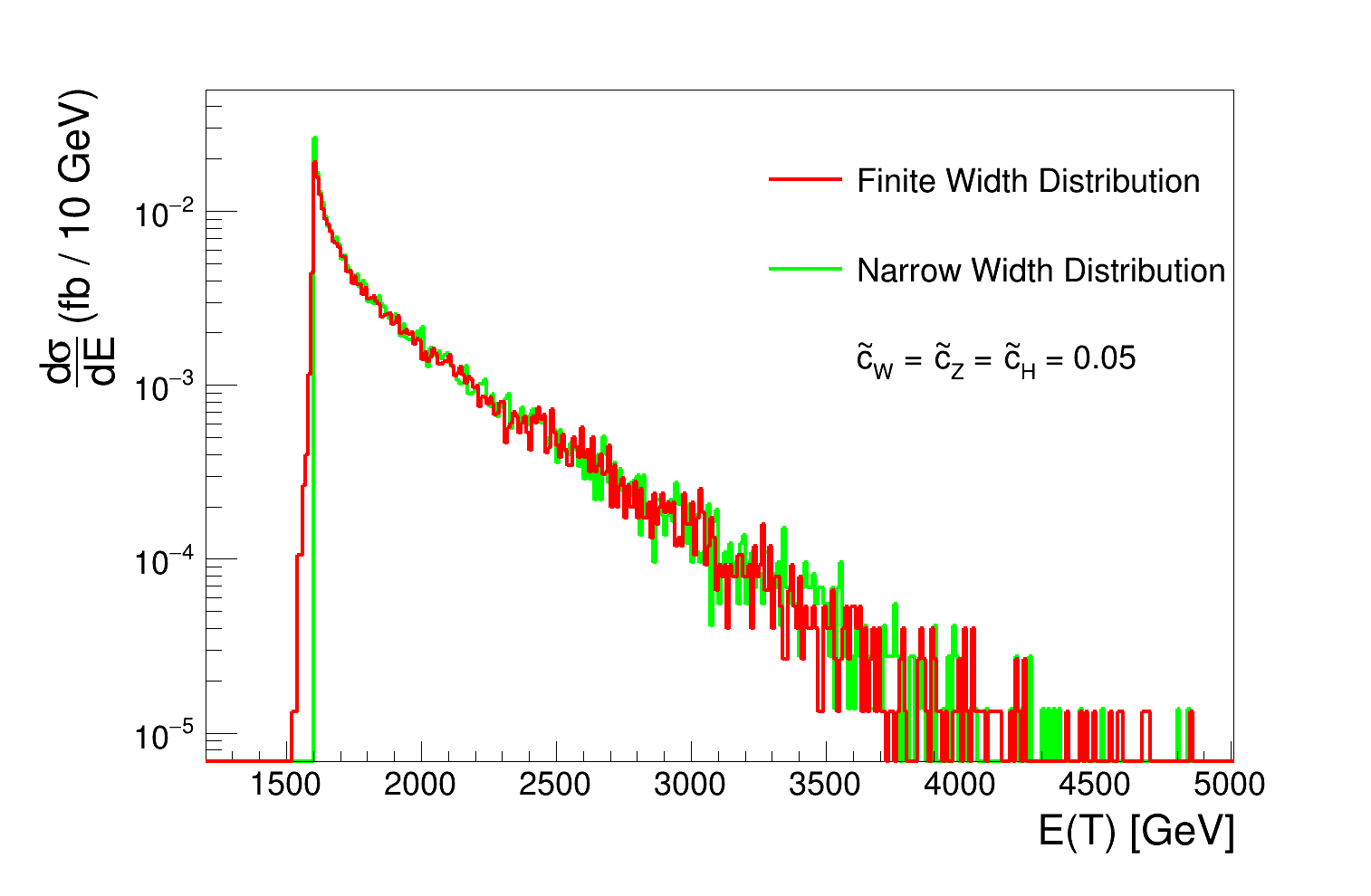}
  \label{fig:ETCompH_005}
  }
  \subfloat[]{
  \includegraphics[width=0.33\textwidth]{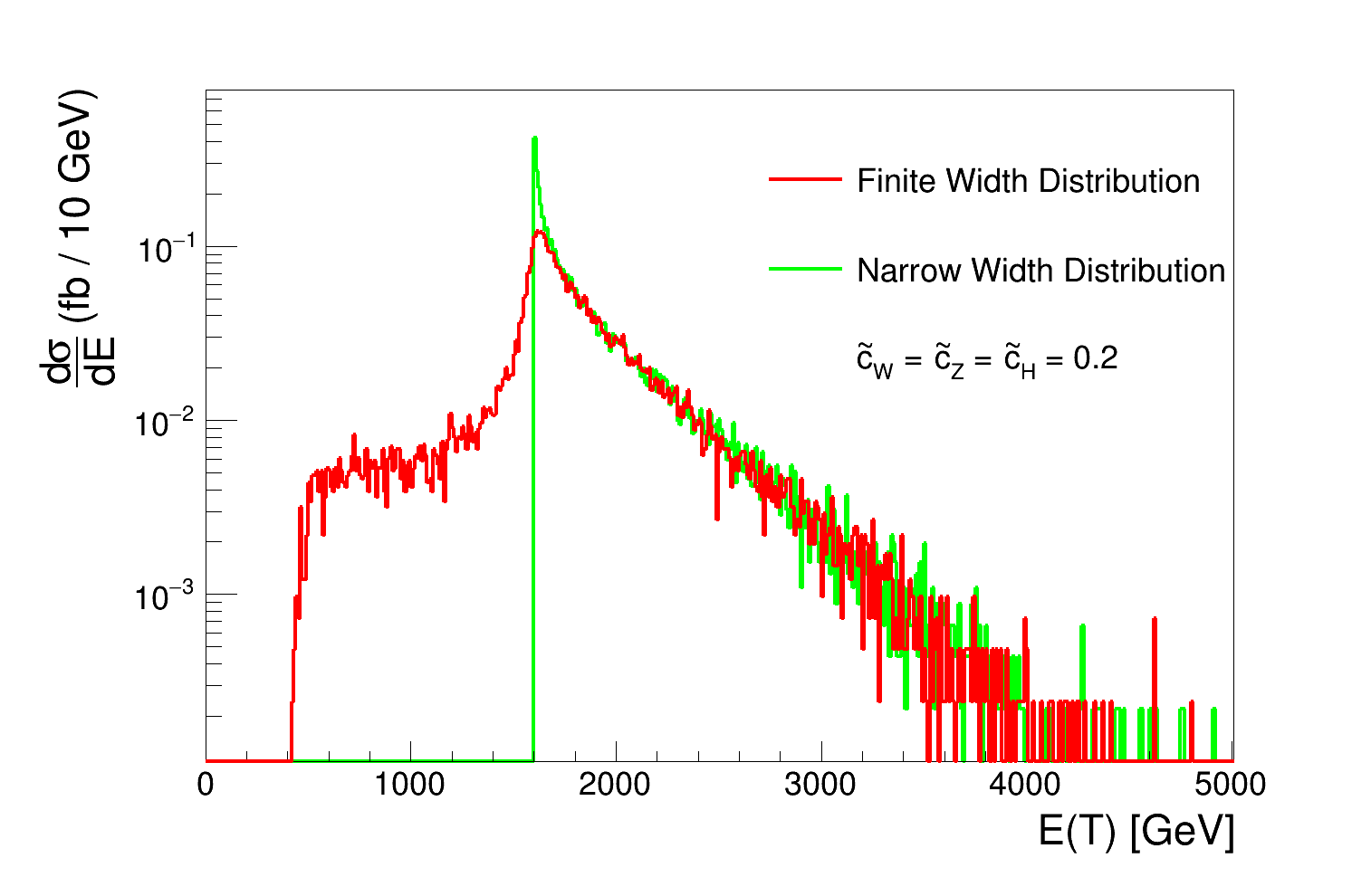}
  \label{fig:ETCompH_020}
  }
  \subfloat[]{
  \includegraphics[width=0.33\textwidth]{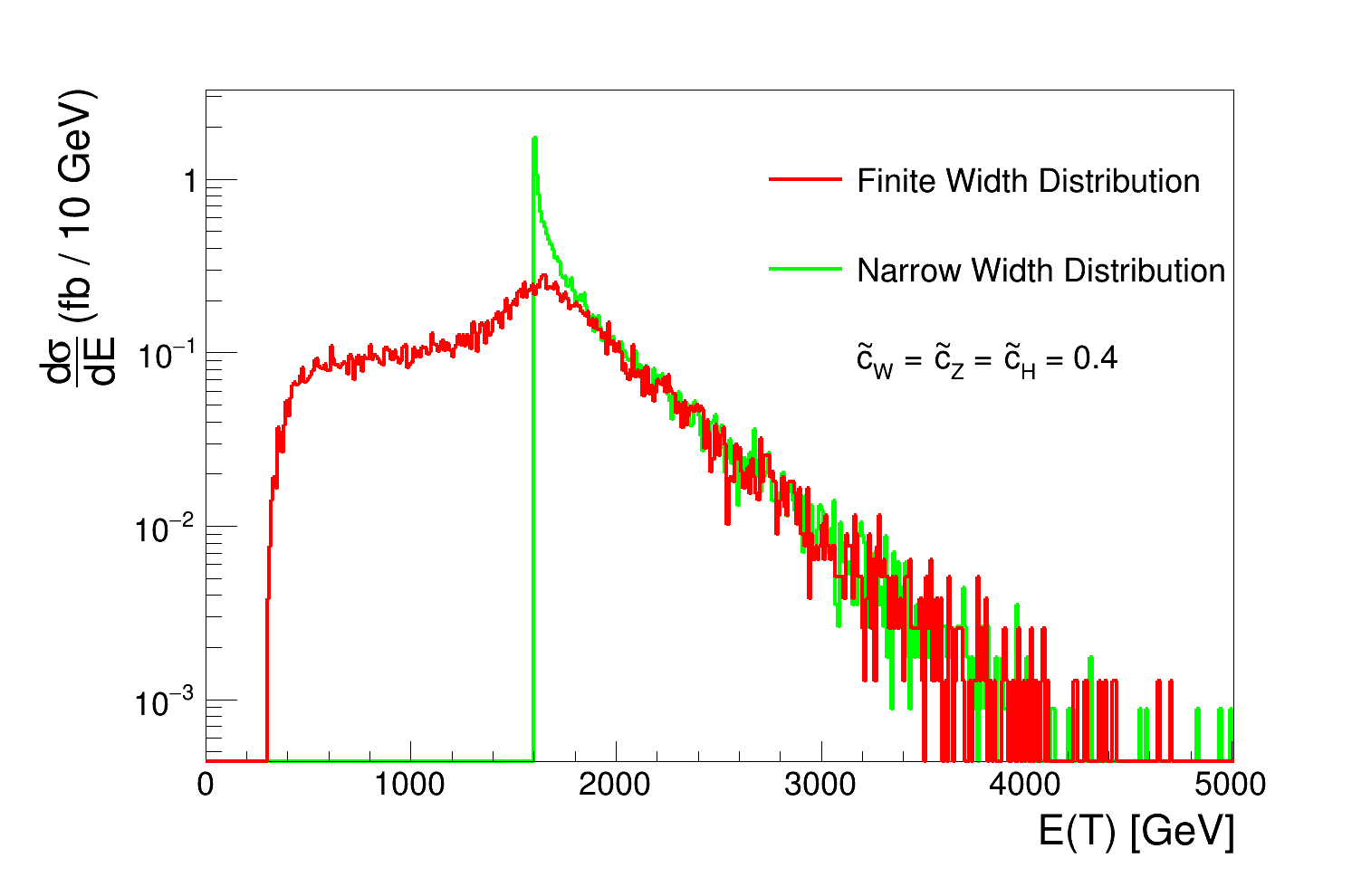}
  \label{fig:ETCompH_040}
  }
  \caption[]{Distribution of differential cross-section of singly produced Top partner ($T$) of mass $M_T = 1.6$~TeV at narrow-width and finite-width as obtained from \textsc{MadGraph\_aMC@NLO} simulation of $WTHt$ process. The coupling values $\cW$ is set 0.05, 0.20, and 0.40 respectively in \subref{fig:ETCompZ_005}, \subref{fig:ETCompZ_020}, and \subref{fig:ETCompZ_040}. All distributions assume $\cW = \cZ = \cH$.}
  \label{fig:ETCompH}
 \end{figure*}

On the other hand, as can be seen in Figure \ref{fig:NWA-Gamma-WTHt}, $\textrm{P}_{\textrm{NWA}}$ becomes slightly higher than unity in the small but finite $\frac{\Gamma}{M}$ region before it starts to decrease. The differential cross-section distributions for $WTHt$ processes of a 1.6~TeV Top partner for different choices of the VLQ couplings are shown in Figure \ref{fig:ETCompH}. At very low decay widths, the finite width cross-section of $WTHt$ processes is slightly smaller than what is predicted by NWA because of widening of the Breit-Wigner propagator, resulting in $\textrm{P}_{\textrm{NWA}} > 1$. However, the VLQ energy distribution for a $WTHt$ process receives an enhancement at lower energies for larger decay widths. This can cause the finite width cross-section for $WTHt$ processes to be higher than what NWA predicts, resulting in $\textrm{P}_{\textrm{NWA}} < 1$.

The functional behavior of $\textrm{P}_{\textrm{NWA}}$ can be approximated by a piece-wise function of the form in equation (\ref{eqn:PNWA-VQHq}).
\begin{widetext}
\begin{align}
%\nonumber
 \textrm{P}_{\textrm{NWA}, VQHq}\left(\frac{\Gamma_Q}{M_Q}\right) &= 
\begin{cases}
1 + AB\frac{\Gamma_Q}{M_Q} - \left(\frac{AB}{x_0}\right) \left(\frac{\Gamma_Q}{M_Q}\right)^2, &  \frac{\Gamma_Q}{M_Q} < x_0 \\
1 - A\left(1 - \exp\left(-B\left(\frac{\Gamma_Q}{M_Q} - x_0\right)\right)\right), &  \frac{\Gamma_Q}{M_Q} \ge x_0
\end{cases}.
\label{eqn:PNWA-VQHq}
\end{align}
\end{widetext}
%&= \left[1 + AB\frac{\Gamma_Q}{M_Q} - \left(\frac{AB}{x_0}\right) \left(\frac{\Gamma_Q}{M_Q}\right)^2\right]\Theta\left(x_0 - \frac{\Gamma}{M}\right) + \left[1 - A\left(1 - e^{-B\frac{\Gamma_Q}{M_Q}}\right)\right]\Theta\left(\frac{\Gamma}{M}-x_0\right)  
%\end{align}
%\end{widetext}
%where $\Theta(x)$ represents the Heaviside step function. 
The two parts of the function are so chosen that $\textrm{P}_{\textrm{NWA}} (x_0) = 1$ and both functions and their derivatives are continuous at the joining point $x_0$. The values of $A, B,$ and $x_0$ parameters for different processes with Top and Bottom partners decaying to the Higgs boson, subject to the constraints $B > 0 $ and $0 < x_0 \le 0.1$, are obtained by a least squared error fit to the observed values of the correction factor from simulation and and tabulated in Table \ref{tab:AB-TB}.

\begin{table*}[]
\centering
 \begin{tabular}{|c | c | c | c | c | c | c | c | c | c | c | c | c |} 
 \hline
   & \multicolumn{3}{|c|}{$WTHt$} 
   & \multicolumn{3}{|c|}{$ZTHt$} 
%   & \multicolumn{2}{|c|}{$WBHb$} 
   & \multicolumn{3}{|c|}{$WBHb$} 
   & \multicolumn{3}{|c|}{$ZBHb$} 
 %  & \multicolumn{2}{|c|}{$ZBHb$} 
 \\
  \hline
 $M_T$ (TeV) & $A$ & $B$ & $x_0$
 			 & $A$ & $B$ & $x_0$
 			 & $A$ & $B$ & $x_0$ 
 			 & $A$ & $B$ & $x_0$

 \\ 
 \hline\hline
1.0 & 0.057 & 35.032 & 0.052
	& $\sim 0$ & $\sim 0$ & -
	\footnote{$\textrm{P}_{\textrm{NWA}, ZTHt} \approx 1.0$ for all $\frac{\Gamma}{M}$ at $M_T = 1.0$~TeV}
%	& 0.047 & 5.136 
	& 0.031 & 96.784 & 0.055
	& 0.162 & 10.297 & 0.042
%	& 0.231 & 3.612
\\
1.2 & 0.189 & 10.005 & 0.035
	& 0.078 & 20.616 & 0.047
%	& 0.169 & 5.046
	& 0.141 & 11.510 & 0.042
	& 0.296 & 7.985 & 0.031
%	& 0.347 & 4.512
\\
1.4 & 0.319 & 7.923 & 0.026
	& 0.199 & 9.290 & 0.030
%	& 0.305 & 4.724
	& 0.264 & 8.116 & 0.030
	& 0.412 & 7.673 & 0.026
%	& 0.465 & 4.870
\\
1.6 & 0.433 & 7.801 & 0.022
	& 0.321 & 7.936 & 0.026
%	& 0.417 & 5.098 
	& 0.379 & 7.633 & 0.025
	& 0.514 & 7.990 & 0.022
%	& 0.555 & 5.597
\\
1.8 & 0.522 & 8.771 & 0.023
	& 0.428 & 8.072 & 0.023
%	& 0.519 & 5.565 
	& 0.476 & 8.204 & 0.023
	& 0.588 & 9.407 & 0.023
%	& 0.638 & 6.266
\\
2.0 & 0.613 & 9.329 & 0.021
	& 0.526 & 8.456 & 0.021
%	& 0.606 & 6.093 
	& 0.568 & 8.773 & 0.023
	& 0.668 & 10.275 & 0.023
%	& 0.707 & 6.960
\\
2.2 & 0.658 & 12.082 & 0.024
	& 0.601 & 9.896 & 0.022
%	& 0.679 & 6.955 
	& 0.633 & 10.563 & 0.023
	& 0.705 & 13.737 & 0.025
%	& 0.763 & 8.004
\\
\hline
\end{tabular}
\caption{The best fit values for the parametric representation of $\textrm{P}_{\textrm{NWA}}$ in equation (\ref{eqn:PNWA-VQHq}) for different values of $M_{T/B}$.}
\label{tab:AB-TB}
\end{table*}

%}

%{
%\color{teal}
% The functional behavior of  $\textrm{P}_{\textrm{NWA}}$ is quite well approximated by a quadratic polynomial for processes that incorporate a decay to the vector bosons. However, for processes where the $T (B)$ decays to a Higgs boson and a top (bottom) quark, the dependence of $\textrm{P}_{\textrm{NWA}}$ takes the form of an exponential decay, well approximated by a function of the form \mbox{$1 - A\left(1-\exp\left(-B\frac{\Gamma}{M}\right)\right)$}. To predict the functional form of the correction factor for the narrow width approximation, we fitted a least squared error regression model for each value of $M_Q$ for all the processes to find the best representative values of the parameters. The best fit values for $A_1$ and $A_2$ ($A$ and $B$ for processes entailing decay to a Higgs boson) for processes involving $T$ and $B$ VLQs are tabulated in Tables \ref{tab:A0A1-T} and \ref{tab:A0A1-B}, respectively.   
%
%
%To summarize, the parametric representation of the $\textrm{P}_{\textrm{NWA}}$ for singly produced VLQ ($Q$) can be given by:
%\begin{widetext}
%\begin{equation}\label{eqn:NWA-1}
%\textrm{P}_{\textrm{NWA}}\left(\frac{\Gamma_Q}{M_Q}\right) = 
%\begin{cases}
%1 + A_1 \frac{\Gamma_Q}{M_Q} + A_2 \left(\frac{\Gamma_Q}{M_Q}\right)^2, &  Q \rightarrow Vq \\
%1 - A\left(1 - \exp\left(-B\frac{\Gamma_Q}{M_Q}\right)\right), &  Q \rightarrow Hq
%\end{cases}
%.
%\end{equation}
%\end{widetext}
%
%}
\section{Reinterpretation of Limits from Existing Analyses}\label{recast}

The proposed interpretation strategy in equation (\ref{eqn:SP-interp-1}) allows a more comprehensive representation of the search results that are currently ongoing in the {ATLAS} and {CMS} experiments. In order to illustrate the flexibility this interpretation strategy offers, we take the ATLAS analysis in~\cite{ATLAS-TZt-run2} and the CMS analysis in~\cite{CMS-TZt-run2} as examples. Both analyses target the search for singly produced top partners $(T_{+\frac{2}{3}})$ that eventually decay to a $Z$ boson, decaying into a pair of electrons or muons, and a top quark. The ATLAS search focuses on two orthogonal analysis channels-- the boosted dilepton channel and the trilepton channel. The boosted dilepton channel looks for a $Z$ boson decaying into a pair of electrons or muons as well as a boosted jet identified as the hadronic shower of the top quark. The trilepton channel includes an additional electron or muon from the leptonic decay of the $W$ boson, emerging from the decay of the top quark produced together with the $Z$ boson. This analysis performs a statistical combination of the two channels and the ATLAS collaboration reports the exclusion limit on the $WTZt$ process cross-section for masses in the range of \mbox{0.7--2.0 TeV} and coupling values, $\kappa$, between 0.1--1.6 following the parameterization prescribed in~\cite{Panizzi}. 
The aforementioned analysis has been stored as a entry in the HEPData~\cite{hepdata} repository.
The exclusion limits on the $WTZt$ process cross-section, as a function of $M_T$ and $\kappa$ are available in this HEPData entry~\cite{hepdata-1}. 
On the other hand, the CMS analysis introduces a ten-category search strategy, based on the combination of lepton flavor from the $Z$ boson decay and the resolution of the $t$ quark decay products. In addition to calculating statistical limits on the $WTZt$  process cross-section under the NWA, assuming a coupling $c_W = 0.5$ and $\BR{T}{Wb} = 0.5$, $\BR{T}{Zt} = \BR{T}{Ht} = 0.25$ for \mbox{0.7 TeV $ \le M_T \le $ 1.7 TeV}, the CMS collaboration also report finite-width cross-section limits for \mbox{$\frac{\Gamma_T}{M_T} = 10\%, 20\%, \textrm{ and} 30\%$} and \mbox{0.8 TeV $ \le M_T \le $ 1.6 TeV}. However, instead of probing the variation of the analysis results in the coupling space, they report their exclusion limit, $\sigma_{\textrm{lim}}$ as a function of relative decay width $\frac{\Gamma_T}{M_T}$ and the top partner mass, $M_T$.  

%{
%\color{red}
Both of these analyses have reported their limits to be chirality-agnostic, only depending on the effective coupling strength of the VLQ couplings and not their chiral structure. The smallest VLQ mass considered in these analyses is 700~GeV, which is large enough that the approximations made in Section \ref{framework} can be applied. For instance, the relative contribution in decay width $\Gamma(T \rightarrow Zt)$ by the interference term between left and right chiral couplings is \mbox{$6\times\frac{\tilde{c}_{L,Z}\tilde{c}_{R,Z}}{\tilde{c}_{L,Z}^2 + \tilde{c}_{R,Z}^2}\times\left(\frac{m_Z}{M_T}\right)^2\frac{m_t}{M_T}$}~\cite{Wulzer}, which evaluates to a maximum value  of 0.013 at $M_T = 700$~GeV. This suggests that the assumptions made in Section \ref{framework} apply to both of the analyses reported in \cite{ATLAS-TZt-run2,CMS-TZt-run2} and we can apply the proposed semi-analytical framework to reinterpret their results.
%}

%Lims_ATLASstyle_ATLAS_0.png
%New_Exclusion_1_ATLAS.png

\begin{figure*}
\centering
  \subfloat[]{
  \includegraphics[width=0.5\textwidth]{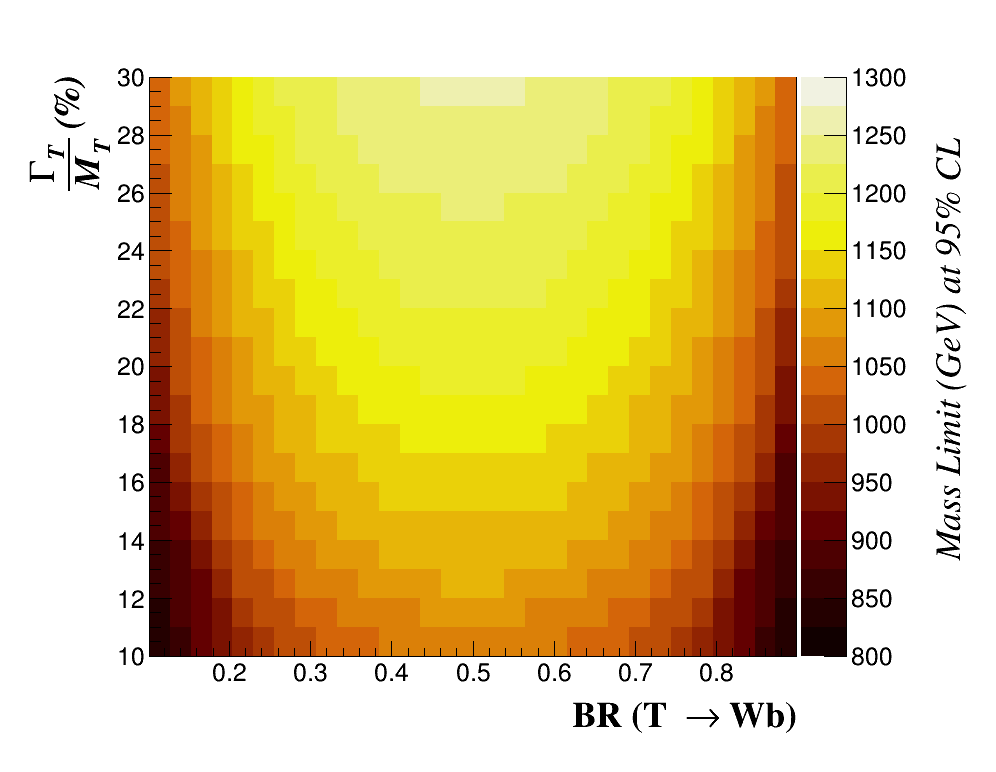}
  \label{fig:new-exclusion-CMS}
  }
  \subfloat[]{
  \includegraphics[width=0.5\textwidth]{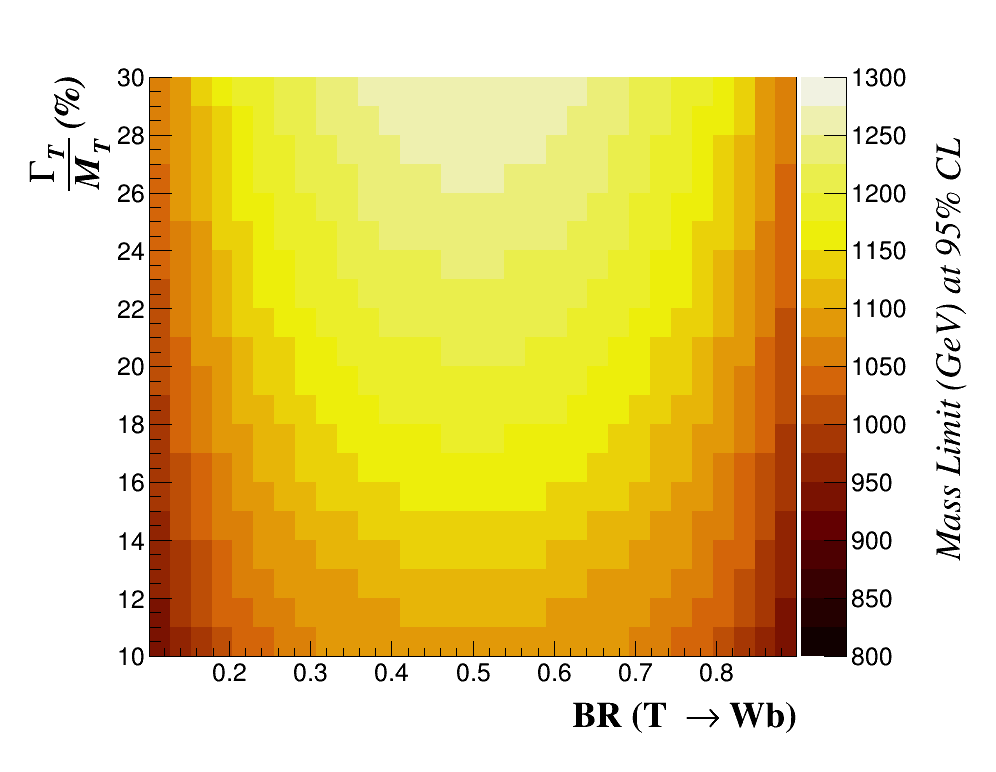}
  \label{fig:new-exclusion-ATLAS}
  }
  \caption[]{Representation of the exclusion limits on VLQ mass in the $\frac{\Gamma_T}{M_T} - \BR{T}{Wb}$ plane. This representation makes the assumption $\cZ = \cH$, deeming the branching ratios in the $H$ and $Z$ channels being equal in the large $M_T$ limit. The limits reported in the CMS analysis~\cite{CMS-TZt-run2} are shown in \subref{fig:new-exclusion-CMS} and the limits obtained in the ATLAS analysis~\cite{ATLAS-TZt-run2} are represented in \subref{fig:new-exclusion-ATLAS}.
} 
  \label{fig:limits-newstyle}
\end{figure*}

Figure \ref{fig:new-exclusion-CMS} gives a generalized representation of the limits reported in the CMS analysis~\cite{CMS-TZt-run2}. To ensure a simplified visualization, we have set \mbox{$\cH = \cZ$} which allows \mbox{$\BR{T}{Zt} \approx \BR{T}{Ht}$} in the large mass limit.  
%The ATLAS analysis in~\cite{ATLAS-TZt-run2} has used a perliminary version of the interpretation strategy laid out in this paper. By making a straightforward use of the inequality in \ref{eqn:SP-interp-1}, they calculate the largest top partner mass excluded at $95\%$ CL for a given choice of couplings. In order to accommodate a straightforward visualization of these limits, the analyzers make an additional simplifying assumption, $\BR{T}{Zt} \approx \BR{T}{Ht}$ which is equivalent to setting $\cH = \cZ$ . 
This assumption is well-motivated in light of the Goldstone Equivalence Theorem~\cite{Goldstone}; in the large-$M_Q$-limit, the longitudinal polarization dominates the $Z$-boson-mediated decay of the top partner which is related to the Higgs mode by a hypercharge rotation, independent of the $SU(2)$ representation of the VLQs~\cite{CompHiggs-1, Aguilar}. As a result, the $Z$ and $H$ boson decay vertices receive similar coupling strengths and the partial decay widths become similar, resulting in almost equal branching ratios independently of the group representation. An equivalent representation of the limits reported by the ATLAS analysis~\cite{ATLAS-TZt-run2} is given in Figure \ref{fig:new-exclusion-ATLAS}.

However, it should be emphasized that the assumption of $\cH = \cZ$, albeit well motivated, is only necessary for the purpose of a convenient representation. We can perform a four-dimensional interpretation by allowing a generalized strategy of parametric reduction. We introduce the $f$ parameter:

\begin{equation}\label{eqn:f-parameter}
f = \frac{\cH}{\cZ}
\end{equation}
which defines a plane of projection in the four-dimensional hyper-space of (\ref{eqn:SP-interp-1}). This parameterization expresses the branching ratios as a function of $\frac{\cZ}{\cW}$ in the large-$M_Q$-limit.
 
\begin{align}\label{eqn:BRs-f}
\BR{T}{Wb} &\approx \frac{1}{1 + (1 + f^2)\frac{\cZ^2}{\cW^2}} \nonumber \\
\BR{T}{Zt} &\approx \frac{1 - \BR{T}{Wb}}{1 + f^2} \\
\BR{T}{Ht} &\approx f^2\times\BR{T}{Zt} \nonumber
\end{align}
 
For a given choice of $f$, the contours for constant branching ratios are represented by vertical straight lines in the $\frac{\Gamma_T}{M_T}$--$\BR{T}{Wb}$ plane. At large $M_Q$ limit, the branching ratios often become independent of the VLQ mass as well as the couplings for certain group representations\cite{Aguilar, Panizzi}. Hence, the $f$-factor-based reduction strategy makes it trivial to evaluate the sensitivity of an analysis in model-specific contexts.

\begin{figure*}
 \centering
 \subfloat[]{
  \includegraphics[width=0.5\textwidth]{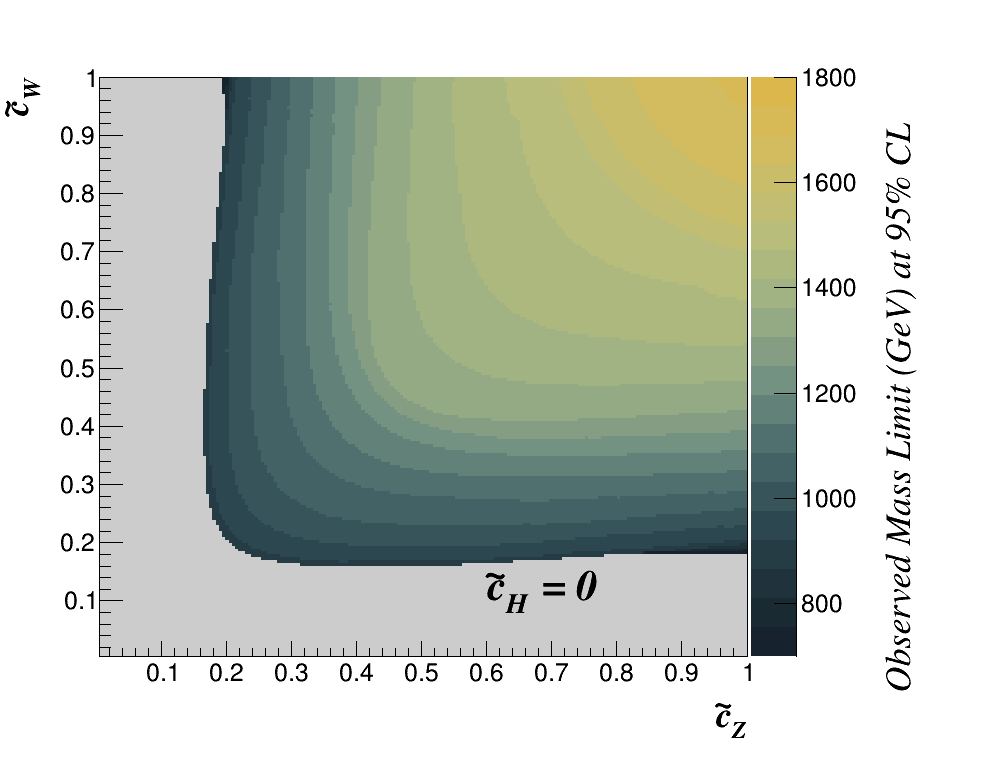}
  \label{fig:limits-ATLASstyle-ch0}
  }
  \subfloat[]{
  \includegraphics[width=0.5\textwidth]{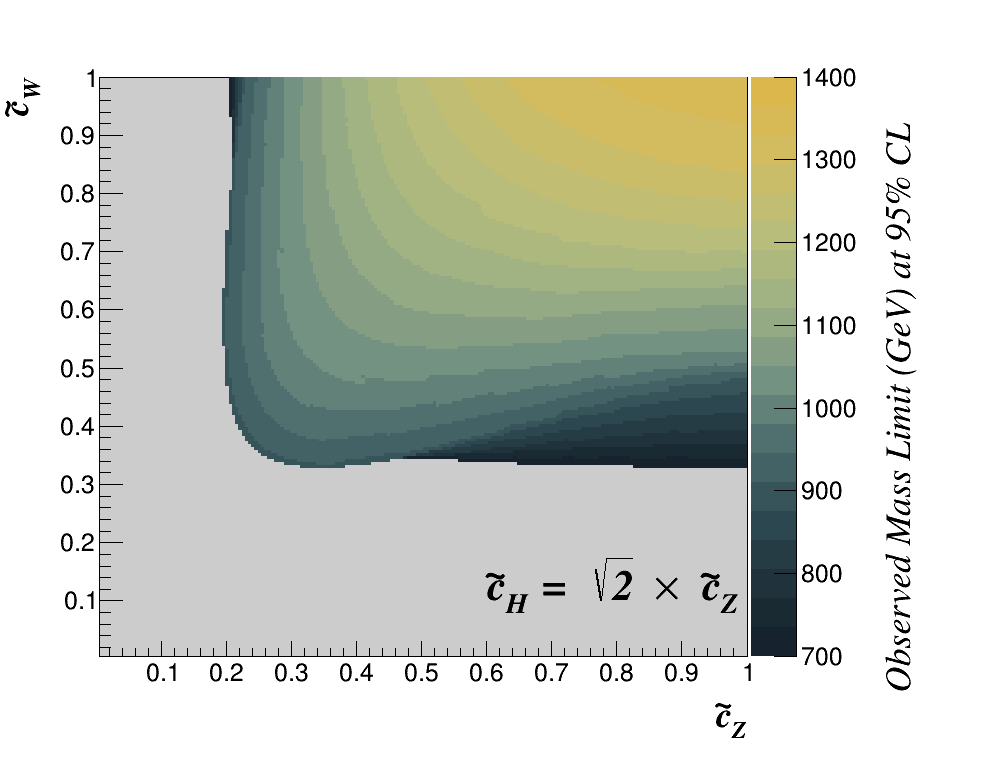}
  \label{fig:limits-ATLASstyle-ch2}
  }
  \caption[]{Reinterpretation of observed limits on top partner mass from~\cite{ATLAS-TZt-run2}, plotted as a function of $\cW$ and $\cZ$ for \subref{fig:limits-ATLASstyle-ch0} $f = 0$ and \subref{fig:limits-ATLASstyle-ch2} $f = \sqrt{2}$. The grey regions enclose a parametric space not covered within the sensitivity of the analysis.}
  \label{fig:limits-ATLASstyle}
 \end{figure*}

  Evidently, the representation of VLQ mass limits as a function of $c_W$ and $c_Z$ in the limit of $\BR{T}{Zt} \approx \BR{T}{Ht}$ proposed in~\cite{ATLAS-TZt-run2} corresponds to the special case of $f = 1$. However, the proposed framework in equation (\ref{eqn:SP-interp-1}) can accommodate other choices of $f$ to probe the exclusion limits on alternate projections of the parametric hyperspace. We illustrate this in Figure \ref{fig:limits-ATLASstyle} where we numerically re-interpret the limits reported by the ATLAS analysis~\citep{ATLAS-TZt-run2} for alternate choices of $f$. In Figure \ref{fig:limits-ATLASstyle-ch0}, we choose $f = 0$ which eventually implies that $\BR{T}{Ht} = 0$. On the other hand, $f = \sqrt{2} \Rightarrow \BR{T}{Ht} \approx 2\times\BR{T}{Zt}$ is chosen for the re-interpretation in Figure \ref{fig:limits-ATLASstyle-ch2}. For a given choice of $f$, the contours of constant branching fractions are given by straight lines passing through the origin in these plots. As expected, the analysis is sensitive to relatively lower top-partner masses for higher values of $f$.  
%
%As expected, the exclusion limits for the choice of $\cH = 0$ are the strongest for all choices of the couplings, excluding masses smaller than 1700 GeV for $\cW \sim \cZ \sim 1.0$ where the corresponding exclusion limit from~\cite{ATLAS-TZt-run2} is around 1500 GeV. On the other hand, the exclusion limit is degraded for $\cH = \sqrt{2}\cZ$, where the $H$ channel has a higher branching ratio compared to the $Z$ channel and hence, reduces the overall sensitivity of the analysis. The exclusion limit is set around 1400 GeV at $\cW \sim \cZ \sim 1.0$.

We now demonstrate how the proposed interpretation strategy can be used to correlate the different strategies used by the aforementioned analyses and hence compare their results. The CMS analysis~\cite{CMS-TZt-run2} adapted the interpretation proposed by Carvalho et. al~\cite{Panizzi-LW}. If a particular analysis is mostly sensitive to a certain decay channel of the VLQ and relatively insensitive to other decay channels, the excluded cross-section, $\sigma_{\textrm{lim}}$ in equation (\ref{eqn:SP-interp-1}), becomes a function of the total VLQ decay width and not the individual choices of the couplings. This is because the change in branching ratio for an alternate choice of couplings that produce the same decay width merely applies as a variation in normalization of the signal hypothesis and hence, is not reflected in the calculation of the exclusion limits for the process cross-section. Hence, the excluded cross-sections themselves can be represented as a function of $M_Q$ and $\frac{\Gamma_Q}{M_Q}$. The exclusion region can be identified by comparing the exclusion limits with the process cross-section which, according to the authors in~\cite{Panizzi-LW}, can be expressed with factorized couplings as given by the following equation:

\begin{equation}\label{eqn:XS-FW}
\sigma_{VQAq} (M_Q, \vec{c})= C_{\textrm{prod}}^2 C_{\textrm{dec}}^2 \times \hat{\sigma}_{VQAq}(M_Q, \Gamma_Q)
\end{equation} 
where $C_{\textrm{prod}}$ and $C_{\textrm{dec}}$ are the couplings associated with the production and decay vertices of the singly-produced VLQs and $\hat{\sigma}$ represents a \textit{reduced cross-section}, that only depends on the choice of VLQ mass and the total decay width. Using the parameterization proposed in sections \ref{framework} and \ref{PNWA}, we can express the so-called reduced cross-section $\hat{\sigma}$ in terms of $\textrm{P}_\textrm{NWA}$ and $\sigma_{\textrm{prod, }WT}^{\textrm{NW}}$ for the $WTZt$ process considered in the analyses by the equation:

\begin{equation}\label{eqn:XS-reduced}
\hat{\sigma}_{WTZt}(M_T, \Gamma_T) \approx \frac{M_T^2 \rho_Z(T)}{8\pi g^2 m_Z^2} \times \frac{\sigma_{\textrm{prod, } WT}^{\textrm{NW}}(M_T, \cW = 1)}{ \frac{\Gamma_T}{M_T} + A_1 \frac{\Gamma_T^2}{M_T^2}  + A_2 \frac{\Gamma_T^3}{M_T^3}} .
\end{equation} 

\begin{table}[]
\centering
 \begin{tabular}{|c | c | c | c|} 
 \hline
 $M_T$ (TeV) & $\frac{\Gamma_T}{M_T}$ (\%) & $\hat{\sigma}$ [pb] from (\ref{eqn:XS-reduced}) & $\hat{\sigma}$ [pb] from~\cite{CMS-TZt-run2}  \\ 
 \hline\hline
 & 10 & 192 & 183  \\
 1.0 & 20 & 92 & 87\\
 & 30 & 59 & 55 \\
 \hline
 & 10 & 141 & 145 \\
 1.2 & 20 & 67 & 68 \\
 & 30 & 42 & 43 \\
 \hline
 & 10 & 107 & 112 \\
 1.4 & 20 & 50 & 52 \\
 & 30 & 31 & 33 \\
 \hline
 & 10 & 80 & 85  \\
 1.6 & 20 & 37 & 39 \\
 & 30 & 23 & 29\\
\hline
\end{tabular}
\caption{Comparison of reduced cross-section values calculated from equation~(\ref{eqn:XS-reduced}) with the values reported in~\cite{CMS-TZt-run2}. The values show good agreement, the difference being at most of $  \mathcal{O}(10\%)$ for most cases.}
\label{tab:red-XS-comp}
\end{table}

\begin{figure}
\centering
  \includegraphics[width=0.5\textwidth]{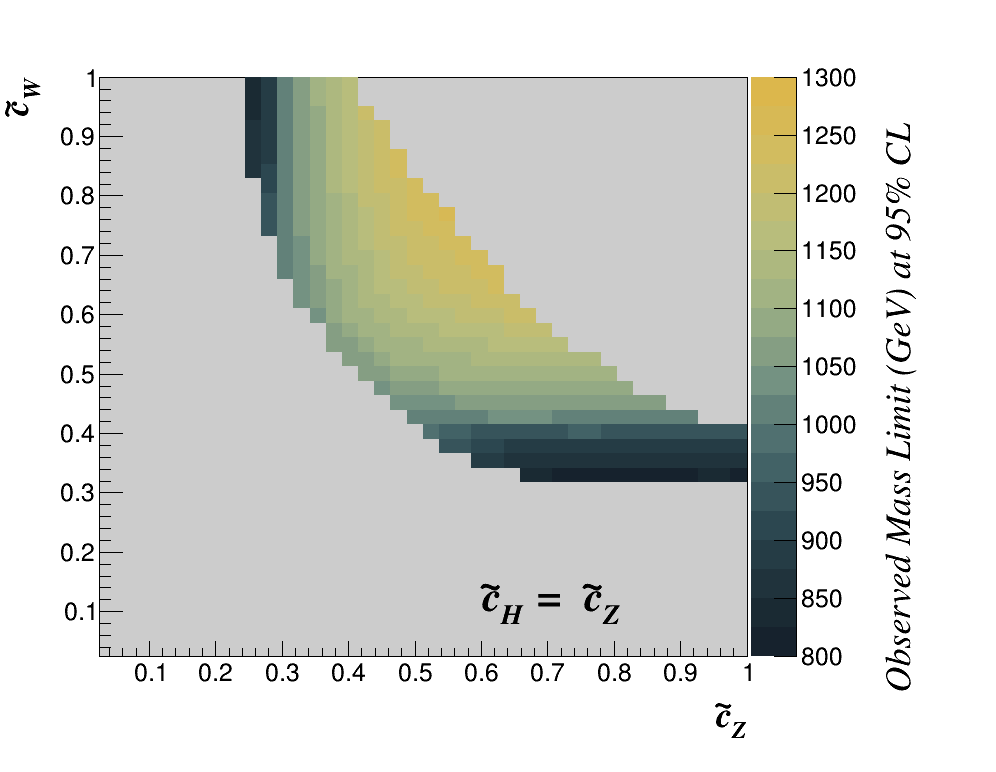}
  \caption{Representation of the exclusion limits on VLQ mass from the CMS analysis~\cite{CMS-TZt-run2} in the \mbox{$\cW$--$\cZ$} plane. This representation makes an assumption $f = 1$. The grey regions enclose a parametric space not covered within the sensitivity of the analysis. }
  \label{fig:limits-CMS2ATLASstyle}
\end{figure}

As shown in Table \ref{tab:red-XS-comp}, equation (\ref{eqn:XS-reduced}) can faithfully predict the reduced cross-section that lies at the heart of the interpretation strategy proposed in~\cite{Panizzi-LW}. This allows us to recast the limits reported in the CMS analysis~\cite{CMS-TZt-run2} as exclusion limits on top-partner mass as a function of the rescaled couplings (Figure \ref{fig:limits-CMS2ATLASstyle}).

\begin{figure}
\centering
  \includegraphics[width=0.5\textwidth]{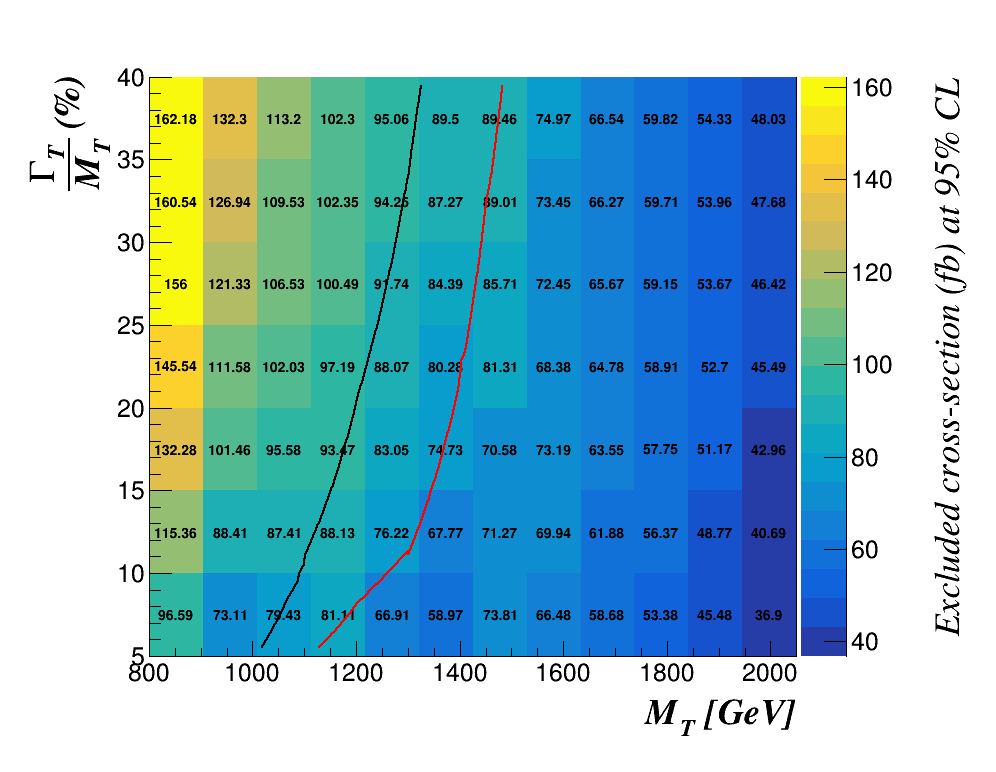}
  \caption{Representation of exclusion limits from the ATLAS analysis~\cite{ATLAS-TZt-run2} in the $\frac{\Gamma_T}{M_T}$--$M_T$ plane. The overlaid black line represents the exclusion limit for the branching fractions to the $W, Z,$ and $H$ boson decay channels set to 0.5, 0.25, and 0.25, respectively. The overlaid red line corresponds to the exclusion limit for branching fractions set to 0.5, 0.5, and 0, respectively. In both cases, the region to the left of the exclusion line is excluded.}
  \label{fig:limits-CMSstyle}
\end{figure}

 As a final example of the flexibility the proposed interpretation strategy offers, we recast the limits from the ATLAS analysis~\cite{ATLAS-TZt-run2} as a function of relative decay width, $\frac{\Gamma_T}{M_T}$, and the top partner mass $M_T$ in figure (\ref{fig:limits-CMSstyle}). In such representations, however, the excluded region in the parametric hyperspace depends on the choice of the branching ratios.

% Finally, we illustrate a different visualization of analysis results based on the currently proposed interpretation strategy. In Figure \ref{fig:limits-newstyle}, the excluded top partner mass limit hass been represented as a function of relative decay width $(\frac{\Gamma}{M_T})$ and the branching ratio in the $W$ mediated decay channel. Similar to the representation in $\cW - \cZ$ plane, this strategy can be expanded to a complete four dimensional interpretation strategy by varying the $f$ factor. 

\section{Conclusion}
We have presented a relatively model-independent approach for interpretation of single VLQ searches. This approach, under a minimal set of assumptions, allows a flexible representation of the results from a VLQ search effort and also provides an avenue of translating results presented in one approach to another. The proposed framework can bridge the gap between experimental searches and their phenomenological reinterpretations in the context of most well motivated VLQ physics models. The novelty of this approach lies in its analytic approach, which makes model-dependent reinterpretations of the search results computationally inexpensive. By numerically recasting the results from two independent analyses, we have established the flexibility the proposed framework offers in obtaining non-trivial, information-dense yet easy-to-interpret representations of such search results. This also harmonizes the representation of single-VLQ search results and hence, provides a platform for the combination of such analyses, an exciting avenue for future work.

\FloatBarrier
%\section*{Acknowledgment}
\begin{acknowledgments}
The work of AR, NN, and TA is supported by the U.S. Department of Energy, Office of High Energy Physics under Grant No. DE-SC0007890. NC acknowledges the support by FCT-Portugal, through project CERN/FIS-PAR/0024/2019.
\end{acknowledgments}
%\newpage
\bibliography{Interpretation_Paper.bib}
\bibliographystyle{apsrev4-1}
\end{document}